\documentclass[12pt]{article}
\usepackage{anyfontsize}
\usepackage{amsfonts}
\usepackage{amsmath,amsthm,amssymb}
\usepackage{mathptmx}
\usepackage{geometry}
\usepackage{graphicx}
\usepackage{subfigure}
\usepackage{hyperref}
\usepackage{cancel}
\usepackage{textcomp}
\usepackage{tikz}
\usepackage{bm}
\usepackage{mathrsfs}
\usepackage{filecontents}
\usepackage{times}
\usepackage{epsfig}
\usepackage{xcolor}
\usepackage{slashed}
\usepackage{booktabs}% build a table
\usepackage{latexsym}
\usepackage{verbatim}
\usepackage{extarrows}
\usepackage{multirow}
\usepackage{rotating}
\usepackage{colortbl}
\usepackage{indentfirst}
\usepackage[numbers,sort&compress]{natbib}
\usepackage{soul}
\definecolor{mygray}{gray}{.9}
\definecolor{intnull}{RGB}{213,229,255}
\usepackage{arydshln}
\usepackage{diagbox}
\usepackage{appendix}
\newcommand{\dif}{\mathrm{d}}

\newcommand{\Hor}{\mathrm{H}}

\usepackage{caption}
\usepackage{tikz}
\usetikzlibrary{arrows,shapes,chains}
\begin{document}
\renewcommand{\thefootnote}{\fnsymbol{footnote}}
\baselineskip=16pt
\pagenumbering{arabic}
\vspace{1.0cm}
\begin{center}
{\Large\sf Scalar fields around a rotating loop quantum gravity black hole: Waveform, quasi-normal modes and superradiance}
\\[10pt]
\vspace{.5 cm}
{{Zhong-Wu Xia${}^{1,}$\footnote{E-mail address: xiazw@mail.nankai.edu.cn}}, 
{Hao Yang${}^{1,}$\footnote{E-mail address: hyang@mail.nankai.edu.cn}},
and
{Yan-Gang Miao${}^{1,2,}$\footnote{Corresponding author. E-mail address: miaoyg@nankai.edu.cn}}
	
	\vspace{6mm}
	${}^{1}${\normalsize \em School of Physics, Nankai University, Weijin Road 94, Tianjin 300071, China}
	
	\vspace{3mm}
	${}^{2}${\normalsize \em Faculty of Physics, University of Vienna, Boltzmanngasse 5, A-1090 Vienna, Austria}
}

\vspace{4.0ex}
\end{center}
\begin{center}
{\bf Abstract}
\end{center}

The rotating loop quantum gravity black hole is a newly proposed non-singular black hole, which eliminates spacetime singularities when a regularization parameter is introduced through loop quantum corrections. This parameter is expected to give rise to observable effects. In this paper, the dynamical behavior of a scalar field near a rotating loop quantum gravity black hole is investigated. Given a small initial perturbation, we obtain the waveform of  massless scalar fields evolving over time.
By analyzing the waveform, we find that the regularization parameter only affects the damping oscillation of waveform,
but not the initial outburst and late-time tail stages. This behavior is  characterized by quasi-normal modes.
Under scalar field perturbations, the loop quantum black holes remain stable.
Moreover, we calculate the quasi-normal modes of massive scalar fields by three numerical methods, which are the Prony, WKB, and shooting methods, respectively. %and compare the accuracy of results among these methods.
%Over the entire parameter space of a rotating loop quantum gravity black hole, 
%Then we analyze  the influence of loop quantum corrections on the quasi-normal modes of scalar field perturbations, 
Our results indicate that the real part of quasi-normal modes depends only on the regularization parameter, while the imaginary part does not only on the regularization parameter but also on the angular momentum.
 Finally, we study the amplification effect of rotating black holes, i.e., the superradiance. 
Our analyses indicate the existence of  stronger superradiance around loop quantum gravity black holes compared to Kerr ones.

\renewcommand{\thefootnote}{\arabic{footnote}}
\newpage
\tableofcontents

%%%%%%%%%%%%%%%%%%%%%%%%%%%%%%%%%%%%%%%%%%%%%%%%%%%%%%%%%%%%%%%%%%%%%%
\newpage
\section{Introduction}
\label{sec:intr}
Although general relativity (GR) is the most widely accepted theory of gravity,  it suffers~\cite{Hawking:1974rv,penrose1969gravitational} from several challenges and unresolved issues, such as singularity, information loss paradox, and breakdown of predictability, etc.
One effective approach to resolve~\cite{Lan:2023cvz} the conundrum of black hole singularities is to construct regular black hole models.
In a regular black hole spacetime,  there are no intrinsic singularities, thus naturally avoiding the issues associated with intrinsic singularities. 
The first regular black hole was constructed~\cite{bardeen1968non} within the scope of GR, but latter many regular black holes have been proposed~\cite{Bonanno:2000ep,Koch:2014cqa,Bouhmadi-Lopez:2020wve} in the framework of modified gravity theories.
%Due to the singularity theorem of Hawking and Penrose \cite{Penrose:1964wq}, it is impossible to construct a regular black hole model satisfying all energy conditions under general relativity.
%The construction of a regular black hole is usually realized in the following ways: 
%\begin{itemize}
    %\item Solving the Einstein field equation under a special symmetry or matter source;
   % \item Directly modifying the metric so that the corresponding space-time has no singularity, and then invert the effective matter term;
    %\item Constructing under the framework of modifying gravity.
%\end{itemize}
Among the various theories of gravity, the loop quantum gravity (LQG) aims to construct a unified theory of quantum gravity to address the issue of spacetime singularities. 
Within the scope of LQG theory, some static and spherically symmetric models of regular black holes have been given~\cite{Modesto:2005zm,Gambini:2013ooa,Bodendorfer:2019nvy,Bodendorfer:2019jay}, where an extra regularization parameter was introduced.% to describe the spacetime of regular black holes.%,BenAchour:2018khr,Ashtekar:2018lag,Blanchette:2020kkk,Alesci:2019pbs,Zhang:2020qxw,Sartini:2020ycs,Assanioussi:2019twp,DeLorenzo:2015taa
%In order to better understand regular black holes, it is necessary to study their observables, such as shadows\cite{Perlick:2021aok}, quasi-normal modes (QNMs)\cite{Konoplya:2011qq,Li:2022kch,Franzin:2022iai}, superradiance\cite{Brito:2015oca,Li:2022kch,Franzin:2022iai} etc.

It is known that astrophysical black holes are rotating, which means that the research on static black holes alone has limited effects on observations.
Recently, a rotating model of regular black holes was suggested~\cite{Brahma:2020eos} in the LQG theory, where its shadow was analyzed and connected to possible future observational data~\cite{Afrin:2022ztr}.
Considering that shadows are only one phenomenon to show the connection between intrinsic properties of black holes and observations, we  
explore other possible phenomena beyond shadows and investigate their potential observable effects  in the scope of LQG theory. 
In the present work we focus on two phenomena: Quasi-normal modes (QNMs) \cite{Konoplya:2011qq,Kokkotas-Schmidt-1999,Li:2022kch,Franzin:2022iai} and superradiance \cite{Brito:2015oca,Li:2022kch,Franzin:2022iai}.

%The ensuing problem is how to detect such regular black holes.
%black holes in nature usually have angular momentum, and the research on the static spherical symmetry model cannot fully meet the detection requirements.
%Recently, a rotating regular black hole model from LQG was proposed, and this study gave a method to judge such black holes through shadows.
%But the parameters of the black hole can only be limited within a certain range through the shadow judgment, and more precise restrictions need to be given by more messenger channels.
%Therefore, this study will start with other important observables different from the shadow, and study the possibility of judging the regular black hole parameters under the quantum gravity of the rotating circle through other messenger signals.
%Here, we focus on two observables, quasi-normal modes and superradiance effects.

Studying perturbations in the background of black holes provides valuable insights into their characteristic signatures, akin to sounds, which can be detected by gravitational wave  detectors. Analyzing matter perturbations, such as scalar field perturbations, holds both theoretical significance and observational relevance within the backdrop of black holes. QNMs of a black hole are characteristic oscillations that arise when the black hole is perturbed. In GR, the QNM frequencies of scalar field perturbations are determined by the mass of scalar fields, the mass and angular momentum of black holes. QNMs are complex due to the existence of event horizons,  so we can divide a QNM frequency into a real part and an imaginary part,
\begin{equation}
    \omega=\omega_{\rm R}+i\omega_{\rm I},
\end{equation} 
where the real part $\omega_{\rm R}$ represents the oscillation frequency and the imaginary part $\omega_{\rm I}$ denotes the decay rate.  
In previous works~\cite{Konoplya:2011qq}, it has been noted that QNMs are highly sensitive to boundary conditions, particularly the asymptotic behaviors of scalar fields near event horizons.  The difference between a LQG metric and a Kerr metric will  lead to differences of boundary conditions and then affect QNMs. In astrophysical observations  the ringdown phase after the merger of two black holes is described by perturbation theory and gravitational waves are a linear superposition of QNMs~\cite{cardoso2016gravitational,gerosa2021hierarchical,Weih:2019xvw}. Through computing the QNMs of scalar field perturbations around a rotating LQG black hole (rLQGBH), we can provide some hints of the underlying gravity theory.

 %QNMs of the scalar field perturbation describe the characteristic oscillations and decay of scalar fields in the vicinity of a black hole. 
 %And these QNMs provide\cite{Cardoso-Pani-2019}  a valuable description of the behavior of scalar perturbations, offering crucial information about the stability, dynamics, and information propagation within black holes. 
%What's more,  the properties of the black hole, such as its mass and angular momentum, as well as  the scalar field itself, determine the QNMs of the scalar field perturbation. 
%By analyzing the frequencies, damping rates, and waveform of these QNMs, we can extract essential details about the black hole's fundamental parameters and investigate\cite{Hod-1998} the underlying gravitational theories. 

%QNM is an observable closely related to the structure of spacetime, which characterizes the relationship of perturbations in spacetime with time. 
%And its specific form is
%\begin{equation}\label{QNM}
%    \omega_{QNM}=\omega_R+i\omega_I
%\end{equation}
%where $\omega_R$ represents the oscillation frequency of the perturbation in spacetime, and $\omega_I$ the decay rate of the perturbation, both of which are determined by the parameters of the spacetime metric.
%Therefore, when the structure of spacetime changes, QNM will change accordingly.

The superradiance effect is  a radiation enhancement process in a dissipative system, such as a near-extreme black hole. In black hole theory, the superradiance is closely associated~\cite{PressTeukolsky1972, Bekenstein1973, Zeldovich1971, StarobinskyChurilov1973} with the ergoregion of rotating black holes.   Especially, the superradiance is a powerful tool to detect \cite{Brito:2015oca,East:2018glu} ultralight scalar fields which are a promising candidate of dark matter. Similar  to QNMs, the superradiance is also sensitive to boundary conditions, thus the regularization parameter %that plays a crucial role in a LQG metric, see Sec.~\ref{sec: Rotating_metric} for the details, 
will leave imprints on the superradiance effects in rLQGBHs, too. We expect to shed some light on the existence of ultralight scalar particles in the LQG theory through the investigation of superradiance. 

Our research focuses on QNMs and superradiance effects, both of which need to deal with Klein-Gordon equations with boundary conditions.
The difficulty of calculations lies in the complicacy of rLQGBHs, where  
one aspect comes from the angular equation due to rotations,  and the other comes from the complicated radial equation.
Although it has widely been applied in GR, see, for instance, Refs.~\cite{leaver1985analytic, leaver1990quasinormal}, the Leaver method is unable to deal with rLQGBHs because a five-term recurrence relation, rather than a three-term one, appears in solving the radial equation of rLQGBHs.
Owing to this reason, we employ three other numerical methods, the Prony, WKB, and shooting methods, to calculate the QNMs of scalar field perturbations around rLQGBHs and compare the results among the three methods in order to obtain the most precise QNMs. Moreover, we analyze the superradiance effect in rLQGBHs by adopting the shooting method which is the most efficient one among the three.
%And the method corresponding to the most accurate results is then applied to calculate the superradiance effect.

The paper is organized as follows.
In Sec.~\ref{sec: Rotating_metric}, we briefly introduce the rotating black holes in loop quantum gravity.
In Sec.~\ref{sec:TD}, we analyze the time domain waveform under scalar field perturbations.
In Sec.~\ref{sec:QNM}, we compute the QNMs of scalar field perturbations around rLQGBHs by using the three numerical methods. 
In Sec.~\ref{sec:superradiance}, we apply the shooting method to calculate the amplification factor. 
Finally, we give our conclusion in Sec.~\ref{sec:con}. The natural units $(G = c = \hbar =1)$ are adopted in our paper.

%%%%%%%%%%%%%%%%%%%%%%%%%%%%%%%%%%%%%%%%%%%%%%%%%%%%%%%%%%%%%%%%%%%%%%
%%%%%%%%%%%%%%%%%%%%%%%%%%%%%%%%%%%%%%%%

\section{Rotating black holes in loop quantum gravity }
\label{sec: Rotating_metric}
In this section we briefly describe the geometry of rLQGBHs.
From a static and spherically symmetric LQGBH, its rotating counterpart was constructed~\cite{Brahma:2020eos,Azreg-Ainou:2014pra} in terms of the modified Newman-Janis algorithm.  
In the Boyer-Lindquist coordinates $(t,r,\theta,\varphi)$, the line element of rLQGBHs reads 
\begin{equation}
\label{1rotating_LQGBH_metric}
{\mathrm d}s^2=-\left(1-\frac{2Mb}{\rho^2}\right) {\mathrm d}t^2
-\frac{4aMb\,{\sin}^2\theta}{\rho^2}{\mathrm d}t{\mathrm d}\varphi
+\rho^2{\mathrm d}\theta^2+\frac{\rho^2}\Delta {\mathrm d}r^2
+\frac{\Sigma\,{\sin}^2\theta}{\rho^2}{\mathrm d}\varphi^2,
\end{equation}
where the relevant quantities are defined by 
%\begin{subequations}
\begin{eqnarray}
\rho^2&=&a^2\mathrm{cos}^2\theta+b^2,\label{eq:rho}\\
%\end{equation}
%\begin{equation}
\Delta&=&b^2+a^2-2Mb, \label{eq:Delta}\\
%\end{equation}
%\begin{equation}
\Sigma&=&\left( b^2+a^2\right) ^2-a^2\Delta\, \mathrm{sin}^2\theta,\label{eq:Sigma}\\
%\end{equation}
%\begin{equation}
b^2&=&\frac{A_\lambda }{\sqrt{1+x^2}}
\frac{M_{\rm B}^2\left( x+\sqrt{1+x^2}\right) ^6+M_{\rm B}^2}{\left( x+\sqrt{1+x^2}\right) ^3},\label{areal}\\
%\end{equation}
%\begin{equation}
M&=&\frac b 2\left[ 1-\frac{8A_\lambda M_{\rm B}^2}{b^2} \left( 1-\sqrt{\frac 1{2A_\lambda}}\frac {1} {\sqrt{1+x^2}}\right)\left(1+x^2\right) \right], \label{Mb}\\
%\end{equation}
%\begin{eqnarray}
x&=&\frac r{\sqrt{8A_\lambda}M_{\rm B}}.\label{eq:x}
\end{eqnarray}
%\end{subequations}
Here the angular momentum $a$ and Arnowitt-Deser-Misner (ADM) mass $M_{\rm B}$ are assumed to be positive, and $A_\lambda$ is a positive dimensionless parameter, here called regularization parameter, which originated~\cite{Bodendorfer:2019nvy} from quantum modifications.
The above rLQGBH is regular~\cite{Brahma:2020eos} everywhere when $A_\lambda>0$ and reduces to a Kerr black hole when $A_\lambda=0$.
By introducing a coordinate  transformation, 
\begin{equation}\label{eq:h}
h=\sqrt{r^2+8A_\lambda M_{\rm B}^2},
\end{equation}
we rewrite $\rho^2$, $\Delta$, $\Sigma$, $b^2$, and $M$ to be 
%\begin{subequations}
\begin{eqnarray}
\rho^2&=&a^2\cos^2\theta+h^2-6A_\lambda M_B^2\\
\Delta&=&h^2+a^2 -2M_{\rm B}h,\label{eq:Delta2}\\
\Sigma&=&\left( h^2-6A_\lambda M_{\rm B}^2+a^2\right) ^2-a^2\Delta \mathrm{sin}^2\theta,\label{eq:Sigma}\\
b^2&=&h^2-6A_\lambda M_{\rm B}^2,\label{eq:b2}\\
M&=&\frac{M_{\rm B}h-3A_\lambda M_{\rm B}^2}{b}.\label{eq:Mb2}
\end{eqnarray}.
   
\begin{figure}[htbp]
		\centering
		\begin{minipage}[t]{0.5\linewidth}
			\centering
			\includegraphics[width=1\linewidth]{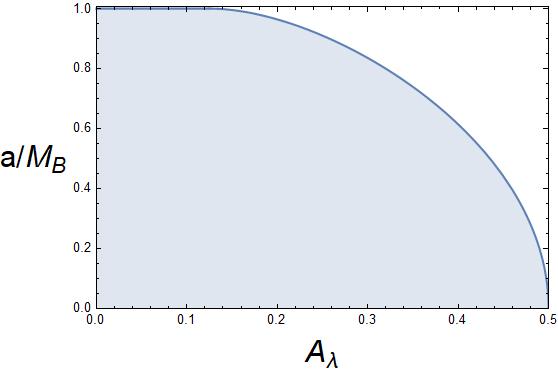}
		\end{minipage}
		\caption{The parameter space $(A_\lambda, a/M_{\rm B})$ of rLQGBH spacetime, where the shadow region is allowed.}
\label{fig:rLQG_parameter}
	\end{figure}

The location of horizons can be determined by the algebraic equation, $\Delta=0$,
and its solutions are
\begin{equation}\label{eq:horizon2}
    	h_\pm=M_{\rm B}\pm \sqrt{M_{\rm B}^2-a^2},
\end{equation}
where a plus or minus sign represents an outer or inner horizon.
According to the existence of an outer  horizon, we restrict the two dimensionless parameters, $A_\lambda$ and $a/M_{\rm B}$ in the shadow region of  Fig.~\ref{fig:rLQG_parameter}, where the blue curve corresponds to the extreme configuration. From the parameter space $(A_\lambda, a/M_{\rm B})$, we obtain that the maximum value of  angular momenta equals to the ADM mass, $a_{\rm Max}/{M_{\rm B}}=1$, which will be adopted in the numerical calculations below. For a given angular momentum $a$, the entire range of regularization parameter $A_\lambda$ is $0\le A_\lambda\le A_{\lambda \mathrm{Max}}$, where
the extreme configuration of rLQGBHs takes the maximum value of $A_{\lambda}$,
\begin{equation}
    A_{\lambda \mathrm{Max}}=\frac{1}{8}\left(1+\sqrt{1-\frac{a^2}{M_{\rm B}^2}}\right)^2.\label{max}
\end{equation}
%When $A_\lambda=A_{\lambda \mathrm{Max}}$, the metric Eq.~(\ref{1rotating_LQGBH_metric}) describes an extreme rLQGBH.

%%%%%%%%%%%%%%%%%%%%%%%%%%%%%%%%%%%%%%%%%%%%%%%%%%%%%%%%%%%%%%%%%%%%%    
\section{Time domain waveform}\label{sec:TD}
%%%%%%%%%%%%%%%%%%%%%%%%%%%%%%%%%%%%%%%%%%%%%%%%%%%%%%%%%%%%%%%%%%%%%
Theoretical studies have shown that the temporal evolution of perturbations in the presence of black holes typically undergoes~\cite{Konoplya:2011qq} three distinct stages. Following the initial pulse, the perturbation field enters a phase of damping oscillations known as quasi-normal ringing, characterized by frequencies and damping times,  i.e. QNMs, solely determined by the black hole parameters. At the last stage of late-time tails, the damping oscillations are overwhelmed by a relaxation process. 

%The evolution of perturbations can be divided~\cite{Konoplya:2011qq} into three distinct stages. At first, the perturbation field undergoes an initial outburst, and then a relatively long period of damping oscillation, and finally an exponential late-time tail. In this section we focus on the impact of quantum parameter $A_\lambda$ on the whole time domain evolution of scalar fields.
%%%%%%%%%%%%%%%%%%%%%%%%%%%%%%%%%%%%%%%%%%%%%%%%%%%%%
\subsection{Numerical method}\label{subsec:setup_TD}
%%%%%%%%%%%%%%%%%%%%%%%%%%%%%%%%%%%%%%%%%%%%%%%%%%%%%
For a massless scalar perturbation around an rLQGBH, the equation of motion is the massless Klein-Gordon equation,
    \begin{equation}\label{eq_KG}
        \nabla^\nu\nabla_\nu\Phi=0,
    \end{equation}
where $\Phi$ is the wave function. %In this section we merely consider massless scalar fileds, so we set $\mu=0$.
Generally, we can simulate the waveform $\Phi$ of scalar fields by solving the above equation in $(t,r,\theta,\varphi)$ coordinates.
 In the Boyer-Lindquist coordinate system, the radial range is taken to be $(h_+, +\infty)$  in computation.  Traditionally, this range is mapped into \((- \infty, +\infty)\) by the tortoise coordinate system. However, it is impossible to perform a numerical integration over a range that extends to \((- \infty, +\infty)\) in practice. Thus, a truncation near the event horizon and  infinity is necessary to render the range finite. Unfortunately, such a truncation introduces~\cite{ThuestadKhannaPrice2017,zhang2020object} artificial boundaries to scalar fields, leading to spurious wave reflections that can spoil the late-time tail of scalar field evolution.
To address the boundary issue inherent in the traditional foliation, a horizon-penetrating hyperbolic (HH) foliation is designed.
%However, the traditional (2+1)-dimensional simulations around rotating black holes often suffer from a serious boundary problem because of the Cauchy foliation, which destroys the precision of numerical calculations. To address this issue, we employ the hyperbolic foliation-dependent strategy~\cite{Zenginoglu2008a,Zenginoglu2008b,Zenginoglu2008c,Zenginoglu2009,Zenginoglu2009b,Zenginoglu2010,Zenginoglu2011a,Zenginoglu2011b} to compute the evolution waveform in time domain in order to avoid the subtle boundary problem through the following two coordinate transformations.
    
In the first transformation, we construct the horizon-penetrating coordinates $\{\tilde t, h,\theta, \tilde \varphi\} $ through 
    \begin{equation}\label{eq:trans_hp}
    	\mathrm d \tilde t=\mathrm d t+ \frac{2Mb}{\epsilon\Delta}\mathrm dh,\qquad
    	\mathrm d \tilde \varphi=\mathrm d \varphi+\frac a {\epsilon\Delta}\mathrm dh,
    \end{equation}
where $\epsilon$ is defined by
\begin{equation}\label{eq:defepsilon}
\epsilon=\sqrt{1-\frac{8A_\lambda M_{\rm B}^2}{h^2}}.
\end{equation}
The rLQGBH metric Eq.~\eqref{1rotating_LQGBH_metric} then becomes 
    \begin{equation}
    	\begin{split}
    		\mathrm ds^2=&-\left(1-\frac{2Mb}{\rho^2}\right)\mathrm d \tilde t^2
    		-\frac{4aMb}{\rho^2}\mathrm{sin}^2\theta\mathrm d\tilde t \mathrm d \tilde \varphi+\frac{4Mb}{\epsilon \rho^2}\mathrm d\tilde t\mathrm d h\\
    		&
    		+\frac 1{\epsilon^2}\left(1+ \frac{2Mb}{\rho^2}\right) \mathrm d h^2
    		-\frac 2\epsilon a\,\mathrm{sin}^2\theta\left( 1+\frac{2Mb}{\rho^2}\right) \mathrm d h \mathrm d \tilde \varphi\\
    		&+\rho^2\mathrm d\theta^2
    		+\left(b^2+a^2+\frac{2Mba^2\,\mathrm{sin}^2\theta}{\rho^2} \right)\mathrm{sin}^2\theta\mathrm d\tilde\varphi^2.
    	\end{split}
    \end{equation}
Since the metric does not contain $\tilde \varphi$ explicitly,  $\partial_{\tilde \varphi}$ is Killing vector. 
Therefore, we can separate variable $\tilde{\varphi}$ from  the scalar field function, 
    \begin{equation}
    	\Phi(\tilde t, h, \theta, \tilde \varphi)=\frac 1h \sum_{m}\psi(\tilde t,h ,\theta)\mathrm e^{im\tilde\varphi},
    \end{equation}
where $m$ is azimuthal number. After substituting the above expression into Eq.~\eqref{eq_KG}, we can express the Klein-Gordon equation as
\begin{equation}\label{eq:KG-hp}
    	A^{\tilde t\tilde t}\partial^2_{\tilde t}\psi
    	+A^{\tilde t h}\partial_{\tilde t}\partial_h\psi
    	+A^{hh}\partial^2_h\psi
    	+A^{\theta\theta}\partial^2_\theta\psi
    	+B^{\tilde t}\partial_{\tilde t}\psi
    	+B^h\partial_h\psi
    	+B^\theta\partial_\theta\psi
    	+C\psi
    	=0,
\end{equation}
where the coefficient functions take the forms,
\begin{eqnarray}
%\begin{split}
A^{\tilde t\tilde t}&=&\rho^2+2Mb,\nonumber\\
A^{\tilde t h}&=&-4\epsilon Mb,\nonumber\\
A^{hh}&=&-\epsilon^2 \Delta,\nonumber\\
A^{\theta\theta}&=&-1,\nonumber\\
B^{\tilde t}&=&2M_B\epsilon-\epsilon\frac{12A_\lambda M_B^2}{h},\nonumber \\
B^h&=&\epsilon^2  \frac 2h(a^2-M_Bh) -2ima\epsilon-\epsilon\Delta \frac{\mathrm d} {\mathrm d h}\epsilon,\nonumber\\
B^\theta&=&-\mathrm{cot}\theta,\nonumber\\
C&=&\frac{m^2}{\mathrm{sin}^2\theta}-\epsilon^2\frac{2(a^2-M_Bh)}{h^2}+\epsilon\frac{2ima}h+\epsilon\frac\Delta h  \frac{\mathrm d} {\mathrm d h}\epsilon.\label{precoefun}
%\end{split}
\end{eqnarray}

In the second transformation, we introduce the hyperbolic foliation~\cite{Harms:2014dqa} to define the compact horizon-penetrating and hyperboloidal coordinates (HH coordinates), $\{\tau, \tilde r,\theta, \tilde \varphi\} $,
    \begin{equation}\label{hyper}
    	\tilde t=\tau+f(\tilde r),\qquad h=\frac {\tilde r}{\Omega(\tilde r)},
    \end{equation}
where 
    \begin{equation}
    	f(\tilde r)=\frac {\tilde r} {\Omega({\tilde r})}-{\tilde r}-4M_{\rm B}\, \mathrm{ln}\,\Omega({\tilde r}),\qquad \Omega({\tilde r})=1-\frac {\tilde r}S.
    \end{equation}
Here $S$ is a constant associated with the hyperbolic foliation. 
According to this coordinate transformation and Eq.~\eqref{eq:horizon2}, we obtain the location of the outer horizon,
\begin{equation}
{\tilde r}_+=\frac{a^2 S+\left(M_{\rm B}+\sqrt{M_{\rm B}^2-a^2}\right)S^2}{a^2+2M_{\rm B}S+S^2},\label{newlocouthor}
\end{equation}
and the relations between the HH coordinates and the horizon-penetrating coordinates,
    \begin{equation}\label{eq:HH-hp}
    	\partial_{\tilde t}=\partial_\tau,\qquad \partial_h=-H\partial_\tau+K\partial_{\tilde r},
    \end{equation}
where 
\begin{equation}\label{eq:defHandP}
H=\frac{\mathrm df({\tilde r})}{\mathrm dh}, \qquad K=\frac{\mathrm d{\tilde r}}{\mathrm dh}.
\end{equation} 
Therefore, the Klein-Gordon equation Eq.~\eqref{eq:KG-hp} can be expressed in the HH coordinates as follows:
   \begin{equation}\label{eq:KG-HH}
   	\partial^2_{\tau}\psi
   	={\tilde A}^{\tau {\tilde r}}\partial_{\tau}\partial_{\tilde r}\psi
   	+{\tilde A}^{{\tilde r}{\tilde r}}\partial^2_{\tilde r}\psi
   	+{\tilde A}^{\theta\theta}\partial^2_\theta\psi
   	+{\tilde B}^{\tau}\partial_{\tau}\psi
   	+{\tilde B}^{\tilde r}\partial_{\tilde r}\psi
   	+{\tilde B}^\theta\partial_\theta\psi
   	+{\tilde C}\psi,
   \end{equation}
where the new coefficient functions have the following connections to the previous ones given by Eq.~\eqref{precoefun},
\begin{eqnarray}
%\begin{split}
\{{\tilde A}^{\tau {\tilde r}},
{\tilde A}^{{\tilde r}{\tilde r}},
{\tilde A}^{\theta\theta},
{\tilde B}^{\tau},
{\tilde B}^{\tilde r},
{\tilde B}^\theta,
{\tilde C}\}
&=&-\frac 1{A^{\tau\tau}}
\{{A}^{\tau {\tilde r}},
{A}^{{\tilde r}{\tilde r}},
{A}^{\theta\theta},
{B}^{\tau},
{B}^{\tilde r},
{B}^\theta,
{C}\},\nonumber\\
A^{\tau\tau}&=&A^{\tilde t\tilde t}-HA^{\tilde t h}+H^2A^{hh},\nonumber\\
A^{\tau {\tilde r}}&=&KA^{\tilde t h}-2KHA^{hh},\nonumber\\
A^{{\tilde r}{\tilde r}}&=&K^2A^{hh},\nonumber\\
B^\tau&=&B^{\tilde t}-HB^h-\frac{\mathrm dH}{\mathrm d {\tilde r}}KA^{hh}, \nonumber\\
B^{\tilde r}&=&K\left(B^h+ \frac{\mathrm dK}{\mathrm d {\tilde r}}A^{hh}\right).\label{HHH}
%\end{split}
\end{eqnarray}

After the above two coordinate transformations, the radial coordinate is mapped to \(\tilde{r} \in [\tilde{r}_+, S)\), where \(\tilde{r}_+\) is defined by Eq.~\eqref{newlocouthor} and \(S \) is an arbitrarily chosen positive constant larger than $\tilde{r}_+$. When executing the numerical calculation, the integration is only performed within the computational region, \(\tilde{r} \in [\tilde{r}_+, S)\). For the other regions, \(\tilde{r} \notin [\tilde{r}_+, S)\), the wave function is set to zero. This mapping confines the range of the radial coordinate to a finite interval, thus naturally imposing boundary conditions without the need for manual specification at the two endpoints. This technique ensures that the scalar field at any moment is only related to itself from the previous moment in \(\tilde{r} \in [\tilde{r}_+, S)\), and unrelated to  the other regions. Such a treatment is equivalent to the boundary conditions with only incoming waves at the event horizon and outgoing waves at the null infinity, naturally coinciding with the boundary conditions for QNMs. And it simplifies the computational region and  avoids the introduction of artificial boundary effects that could affect the accuracy of simulation results for the scalar field's evolution.
Therefore, the HH coordinate system does not require any truncation to perform the numerical integration.
In order to numerically solve Eq.~\eqref{eq:KG-HH} , we introduce an auxiliary function $\Pi$, so that we can reduce this equation to two first-order equations,
%\begin{subequations}
\begin{eqnarray}
\partial_\tau \psi&=&\Pi,\label{RK1}\\
%\end{equation}
%\begin{eqnarray}
%\begin{split}
\partial_\tau \Pi&=&{\tilde B}^{\tau}\Pi
+{\tilde A}^{\tau {\tilde r}}\partial_{\tilde r}\Pi
+{\tilde A}^{{\tilde r}{\tilde r}}\partial^2_{\tilde r}\psi
+{\tilde A}^{\theta\theta}\partial^2_\theta\psi
+{\tilde B}^{\tilde r}\partial_{\tilde r}\psi
+{\tilde B}^\theta\partial_\theta\psi
+{\tilde C}\psi,\label{RK2}
%\end{split}
\end{eqnarray}
%\end{subequations}
which can be solved by the fourth-order Runge-Kutta method~\cite{schiesser2012numerical}. 
Here we take a Gaussian wave packet as the initial condition,
\begin{eqnarray}
%\begin{split}
\psi(\tau=0,{\tilde r},\theta)&=&Y_{lm}(\theta)\,{\exp}\left[-\frac{({\tilde r}-{\tilde r}_c)^2}{2\sigma^2}\right],\nonumber \\
\Pi(\tau=0,{\tilde r},\theta)&=&0,
%\end{split}
\end{eqnarray}
where $Y_{lm}(\theta)$ is the $\theta$-dependent part of spherical harmonics, and $\tilde r_c$ and $\sigma$ are the center and width of Gaussian packets, respectively. We choose the location of our observer to be $\tilde r_c = 6M_{\rm B}$, $\theta=\frac{\pi}{4}$,
 the width of wave packets to be $\sigma=0.2$, and the free parameter to be $S=10$ as suggested by Refs.~\cite{Harms:2014dqa,zhang2020object}, then we solve Eqs.~\eqref{RK1} and \eqref{RK2} and obtain the time domain evolution profiles.
 %{\color{red}With HH coordinate, we do not need to specify boundary conditions any more as the QNM boundary conditions, where only incoming waves exist at the event horizon and only outgoing waves at infinity, are naturally satisfied.}
 
 \begin{figure}[htbp]
 	\centering
 	\begin{minipage}[t]{0.5\linewidth}
 		\centering
 		\includegraphics[width=1\linewidth]{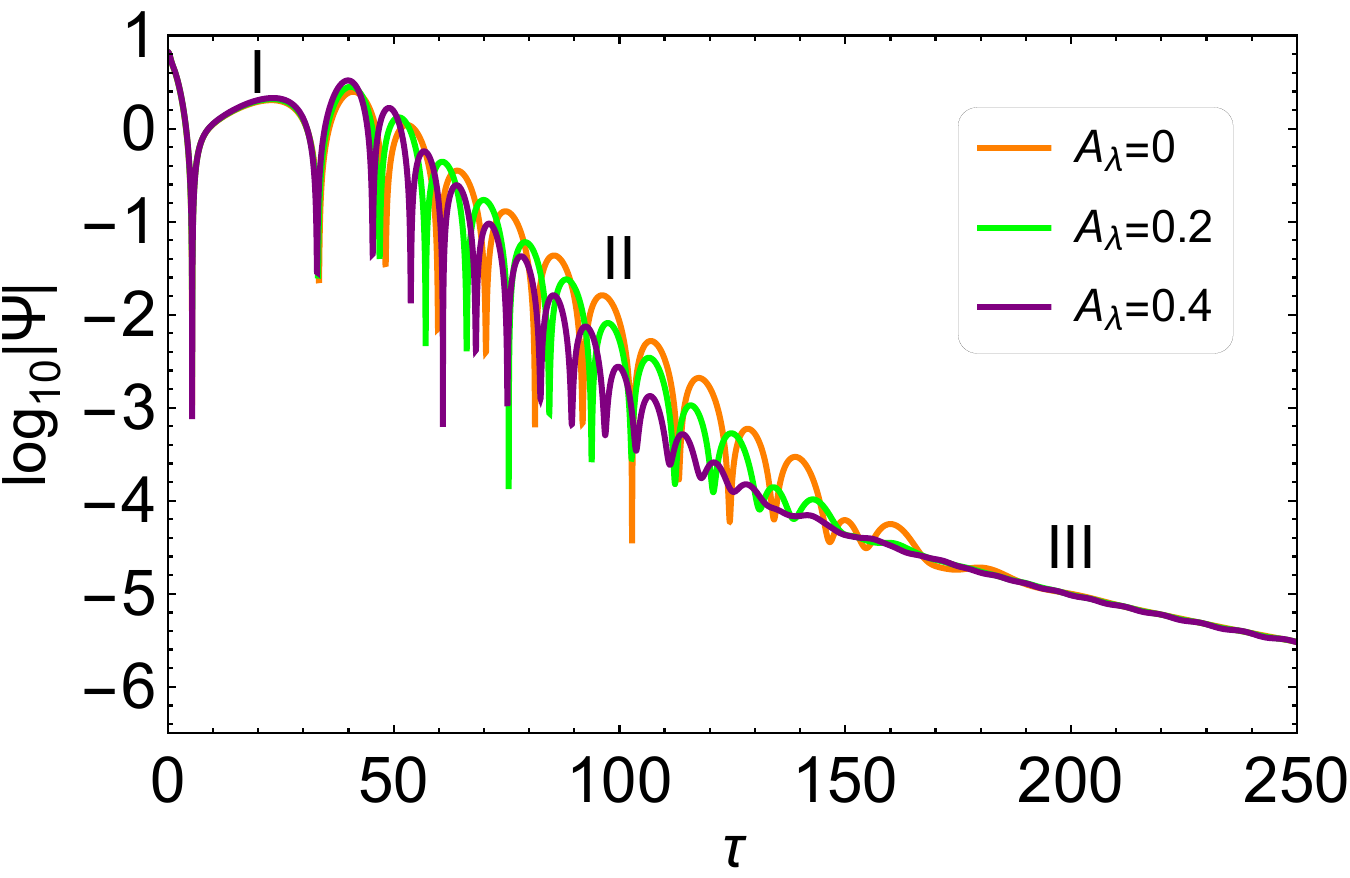}
 	\end{minipage}
 	\caption{
 		The waveform of massless scalar perturbations with various values of  $A_\lambda$, where the initial mode is chosen to be $l=1=m$, and $M_{\rm B}=1$ and $a=0.1$ are set. Three distinct phases are notable, where  phase $\mathrm I$ is initial outburst, phase $\mathrm{II}$ damping oscillation, and phase $\mathrm{III}$ late-time tail.}
 	\label{instab}
 \end{figure}

%%%%%%%%%%%%%%%%%%%%%%%%%%%%%%%%%%%%%%%%%%%%%%%%%%%%%%%%%%%%%%%%%%%%%%
\subsection{Results}\label{sec:TD-result}
%%%%%%%%%%%%%%%%%%%%%%%%%%%%%%%%%%%%%%%%%%%%%%%%%%%%%%%%%%%%%%%%%%%%%%
 %Generally, a rotating black hole exhibits~\cite{zhang2020object} mode-mixing phenomena, that is, an initial even (odd) mode with multipole number $l$ can be excited to other even (odd) modes with the same azimuthal number $m$.
 We choose $l=1=m$ %and $l=2$ respectively 
 as  our initial  mode and draw the waveform of a massless scalar field perturbation around an rLQGBH in Fig.~\ref{instab}.  
 Analyzing the evolution profiles, we observe that the introduction of regularization parameter $A_\lambda$ does not significantly affect the evolution of outbursts and late-time tails. 
 However, it exerts a notable influence on the damping oscillation phase, that is, an increase of $A_\lambda$ leads to an accelerated oscillation and a  rapider decay of scalar fields. Therefore, in order to distinguish LQG from GR, i.e., an rLQGBH from a Kerr black hole, we need to study the QNM frequencies of the damping oscillation stage.
 In the subsequent section, we provide a detailed investigation on the influence of $A_\lambda$ on the  QNMs under scalar field perturbations.

%%%%%%%%%%%%%%%%%%%%%%%%%%%%%%%%%%%%%%%%%%%%%%%%%%%%%%    
\section{Quasi-normal modes}\label{sec:QNM}
%%%%%%%%%%%%%%%%%%%%%%%%%%%%%%%%%%%%%%%%%%%%%%%%%%%%%%
 If a scalar field meets the specific boundary conditions:\footnote[1]{Such boundary conditions always hold for any asymptotically flat spacetime, such as the rLQGBHs we are considering. However, for any asymptotic AdS spacetime, there exist other boundary conditions and corresponding QNMs, see for instance, one of our recent papers~\cite{Guo:2024jhg}.} Pure ingoing waves exist at the outer horizon of black holes, while pure outgoing waves exist at the spatial infinity, the characteristic complex frequencies of damping oscillations are just QNMs.
Owing to the  complicated boundary conditions around rLQGBHs, it is not feasible to obtain QNMs precisely through analytical methods. 
As a result, various numerical and semi-analytical methods have been adopted~\cite{Konoplya:2011qq, Kokkotas-Schmidt-1999, Li:2022kch, Berti:2007dg, Chandrasekhar:1975zza, Molina:2010fb, Franzin:2022iai, Iyer:1986nq, Seidel:1989bp}. 
Each method possesses its own advantages and disadvantages, so that the accuracy of results cannot be guaranteed if one relies solely on a single method. 
In this section, we employ three methods to compute QNMs and make comparisons among them in order to ensure the accuracy and reliability of our findings.

 As shown in Ref.~\cite{zhang2020object}, the choice of coordinate systems mainly affects the late-time tail stage  of scalar field  evolution but hardly the second phase, the damping oscillation stage. Since QNMs are characteristic frequencies of the damping oscillation stage, both the conventional and horizon-penetrating hyperbolic coordinate systems are suitable. Considering that the radial equation in the former system is much simpler than that in the latter one, we choose the former coordinate system when computing QNMs.

%%%%%%%%%%%%%%%%%%%%%%%%%%%%%%%%%%%%%%%%%%%%%%%%%%%%%%%%%%%%%%%%%%%%%%    
\subsection{Prony method}\label{sec:prony}
%%%%%%%%%%%%%%%%%%%%%%%%%%%%%%%%%%%%%%%%%%%%%%%%%%%%%%%%%%%%%%%%%%%%%%
In Sec.~\ref{sec:TD} we have presented the approach for obtaining time domain profiles of massless scalar field perturbations in rLQGBHs. Now we can extract QNMs from these profiles by the Prony method~\cite{Konoplya:2011qq,Berti:2007dg}.     
The main idea of this method is to fit the profile data with a superposition of damping exponents,
\begin{equation}\label{eq:prony}
    \Phi(t)\sim \sum_{j=1}^p C_j e^{-i\omega_jt},
\end{equation}
where $C_j$ is the amplitude coefficient of QNMs. In computations, we extract $(2p-1)$ equidistant points from the damping oscillation phase of a profile. 
These points are used to form one $p\times p$ matrix in order to calculate $\omega_j$. 
Here rows and columns of this metric are composed of equidistant points with decreasing sequence numbers. 
Among the $p$ $\omega_j's$, the one with the largest $|C_j|$ is dominant, which is our demanding QNM frequency.

Although the QNMs obtained through the Prony method exhibit~\cite{zhang2020object} a high precision, our (2+1)-dimensional simulation approach cannot capture the waveform of massive scalar fields, making the QNMs of massive scalar field perturbations unavailable.
Therefore, alternative methods must be employed to compute the QNMs for massive scalar field perturbations in the subsequent sections.%restriction: Since our analysis of time domain waveforms in Sec. \ref{sec:TD} cannot  handle massive scalar fields, we cannot extract QNMs  from corresponding evolution profiles by the Prony method.
%the waveform data in a certain situation is missing, the corresponding QNMs cannot be obtained.For example, the (2+1)-dimensional simulation used  is not capable of handling the case involving a massive scalar field. Therefore we can not obtain the QNMs of a massive scalar field through .
%However, one of our main goals is investigating the ultralight scalar field.

\subsection{WKB method}
The WKB method is a semi-analytic technique for determining low-lying QNMs~\cite{Seidel:1989bp}. 
For a massive scalar perturbation around an rLQGBH, the massive Klein-Gordon equation is
    \begin{equation}\label{eq_m}
        \nabla^\nu\nabla_\nu\Phi=\mu^2 \Phi,
    \end{equation}
where $\mu$ is the mass of scalar fields. Assuming that a scalar field has the same symmetry as that of its background spacetime, we can make the following ansatz,
\begin{equation}\label{eq:Phi-SR}
    	\Phi(t,h,\theta,\varphi)=\int{e^{-i\omega t}\sum_{l,m}S_{lm}(\theta)R_{lm}(h)e^{im\varphi}}\dif\omega,
\end{equation}
where $h$ is defined by Eq.~(\ref{eq:h}), $l$ stands for the angular momentum number, and $\omega$ denotes the characteristic frequency of scalar fields.
After separating variables, we obtain the angular equation,   
\begin{equation}\label{azimuthal}
    	\frac 1 {\mathrm{sin}\theta} \frac {\mathrm d} {{\mathrm d}\theta}\left[ \mathrm{sin}\theta\frac {\mathrm d} {{\mathrm d}\theta}S_{lm}(\theta)\right] 
    	+\left[ a^2\left( \omega^2-\mu^2\right) \mathrm{cos}^2\theta-\frac{m^2}{\mathrm{sin}^2\theta}+A_{lm}\right] S_{lm}(\theta)=0,
    \end{equation}
where $S_{lm}(\theta)$ is spherical harmonics~\cite{Franzin:2022iai}, which reduces to $Y_{lm}(\theta)$ when $a^2(\omega^2-\mu^2)\cos^2\theta=0$, and $A_{lm}$ is angular eigenvalue. The radial equation reads
    \begin{equation}\label{radialh}
    	\epsilon \frac {\mathrm d}{{\mathrm d}h}\left[\epsilon \Delta\frac {\mathrm d}{{\mathrm d}h}R_{lm}(h)\right] +\left[ \frac{\tilde{K}^2}{\Delta}-\mu^2b^2-(A_{lm}-2am\omega+a^2\omega^2) \right] R_{lm}(h)=0,
    \end{equation}
  where $\tilde{K}=am-(b^2+a^2)\omega$.
By introducing the following transformation,
    \begin{equation}\label{eq:transform}
        \Psi(y)=(b^2+a^2)^{1/2}R_{lm}(h),
    \end{equation}
we can change the radial equation of motion  to a Schr\"odinger-like one,
\begin{equation}\label{WKB}
\frac {\mathrm d^2}{{\mathrm d}y^2}\Psi(y) +\left[ \omega^2-V(y,\omega)\right]  \Psi(y)=0,
    \end{equation}
where $y$ is the tortoise coordinate determined by 
    \begin{equation}
        \frac{\mathrm d h}{{\mathrm d}y}=\frac{\epsilon\Delta}{b^2+a^2},
    \end{equation}
and $V(y,\omega)$ is the effective potential,
\begin{eqnarray}
%\begin{split}
V(y,\omega)&=&V_1+V_2,\nonumber \\
V_1&=&\frac{\Delta }{h(b^2+a^2)^4}\Big\{-96 A_\lambda^2 M_{\rm B}^4(M_{\rm B} - h) + a^4 h + 2 M_{\rm B} h^4 \nonumber \\
& &- 2 A_\lambda M_{\rm B}^2 h^2 (4 M_{\rm B} + 5 h) + 
  a^2 \left[h^2 (-4 M_{\rm B} + h) + 2 A_\lambda M_{\rm B}^2(8 M_{\rm B} + h)\right]\Big\},\nonumber \\
V_2&=&\frac{\Delta }{(b^2+a^2)^2}\left[\frac{\tilde{K}^2}{\Delta}-\mu^2b^2-(A_{lm}-2am\omega+a^2\omega^2)\right]. 
%\end{split}
\end{eqnarray}

According to Refs.~\cite{Franzin:2022iai, Konoplya:2019hlu}, we find that the fourth-order WKB approximation is the most suitable for the case of rotating black holes among all orders' WKB approximations.
Therefore, we choose the fourth-order WKB approximation to numerically solve the radial equation. 
Despite its minimal computational resource requirement, the fourth-order WKB approximation exhibits a very small error compared to the more precise sixth-order WKB approximation. 
The QNM frequency $\omega$ is given by the WKB formula \cite{Iyer:1986nq, Konoplya:2003ii}:
    \begin{equation}\label{WKB_formula}
    	\frac{i\left[ \omega^2-V(y_0,\omega)\right] }{\sqrt{-2V''(y_0,\omega)}}-\sum_{j=2}^{4}\Lambda_j=n+\frac 1 2,\qquad n=0,1,2,\cdots,
    \end{equation}
where the prime means the derivative with respect to the tortoise coordinate, and $y_0$ is fixed by the condition, 
\begin{equation}\label{eq:V0}
    \frac{\mathrm d}{\mathrm dy}V(y,\omega)\Big|_{y=y_0}=0,
\end{equation}
and $\Lambda_j$ denotes higher order correction terms.
The specific form of $\Lambda_j$ is given in Ref.~\cite{Konoplya:2003ii}.

Rotating black holes are more intricate than static ones because both the scalar potential $V(y,\omega)$ and the angular eigenvalue $A_{lm}$ depend on the frequency $\omega$ in the case of rotating black holes. For massless scalar field perturbations, we utilize~\cite{Seidel:1989bp} the series expansion method  to solve Eq.~\eqref{WKB_formula}, retaining the terms up to $(a\omega)^6$. However, for massive scalar field perturbations, the angular eigenvalue $A_{lm}$ should be expanded as a series of $(a\sqrt{\omega^2-\mu^2})$ because only even-order terms appear due to the absence of spin for scalar particles. The expansion takes the form:
\begin{equation}
    A_{lm}=\sideset{_{0}}{^{lm}_0}{\mathop f}
    +\sideset{_{0}}{^{lm}_2}{\mathop f}a^2(\omega^2-\mu^2)
    +\sideset{_{0}}{^{lm}_4}{\mathop f}a^4(\omega^2-\mu^2)^2
    +\cdots,
\end{equation}
and $\sideset{_{0}}{^{lm}_j}{\mathop f}$ is the $j$th order expansion coefficient.
By employing the expressions of $A_{lm}$ and $y_0$, we can also express both $V_0$ and its higher order derivatives as series of $(a\omega)$ up to order $(a\omega)^6$. 
Finally, we solve Eq.~\eqref{WKB_formula} numerically and determine the QNMs with the fixed values of $a$, $\mu$, $n$, $l$, and $m$.    

The advantage of the WKB method lies in its polynomial formula with which we can obtain QNMs with high accuracy through simple numerical calculations. 
As discussed above, this method depends only on the first six terms of series expansions. However, these terms diverge in certain cases, which leads to a breakdown of the method. 
The applicable condition can be given in a specific range of parameter $a$. 
In calculations we cannot directly apply the WKB formula to rotating black holes because the radial potential $V(y,\omega)$ in Eq.~\eqref{WKB} involves an unknown frequency. Instead, we seek help from the series expansion. When requiring the convergence of this series, we obtain the applicable condition under an appropriate approximation. For the detailed discussions, see Sec.~\ref{sec:MPC}, where the WKB method is extrapolated to be valid in the range   of $a\lesssim0.4$.         
    
\subsection{Shooting method}\label{shooting}
%The shooting method is a numerical technique \cite{Chandrasekhar:1975zza,Franzin:2022iai} for the calculation of QNMs in black hole physics. 
%Its main idea lies in numerically integrating radial equation of motion from one  point near an event horizon to an intermediate point and also from the other point near the spatial infinity to this intermediate point,  and then we require that both the wave functions and their first derivatives obtained from the two sides are equal at the intermediate point. 

The shooting method is a straightforward numerical strategy~\cite{Chandrasekhar:1975zza,Franzin:2022iai} for solving ordinary differential equation and determining QNMs. The key issue to obtain QNMs is the precise numerical integration of the radial equation of motion. This process initiates from a proximal vicinity of the event horizon and extends toward an arbitrarily selected intermediate point. Conversely, a similar integration procedure is conducted from a position near the spatial infinity back to the same intermediate locale. The crux of the method hinges upon ensuring continuity and differentiability of the wave functions across this junction, that is, both the wave functions and their first-order derivatives, as calculated from the aforementioned disparate starting points, must converge and exhibit equality at the intermediate point. This condition ensures the physical validity and continuity of the solution in the domain from the event horizon to infinity, reflecting the underlying physics of black hole perturbations and the propagation of waves in their vicinity.

%To achieve the goal, we analyze the asymptotic behaviors of the radial equation, Eq.~\eqref{radialh}. The outer event horizon $h=h_+$ and spatial infinity $h=h_\infty$ are two regular singularities, thus we can give the radial wave function $R(h)$ by two  series that are convergent in the range of $h_+<h<h_\infty$.Near the outer horizon, the asymptotic formulation of radial wave function $R(h)$ reads

To fulfill our objectives, we conduct a comprehensive analysis of the asymptotic behaviors associated with the radial equation as delineated in Eq.~\eqref{radialh}. Notably, the outer event horizon, denoted as $h=h_+$, and the spatial infinity, represented by $h=h_\infty$, constitute two regular singular points for the radial equation. Consequently, it becomes feasible to represent the radial wave function $R(h)$ through two convergent series near the two boundaries.

Near the outer event horizon,  the radial wave function $R(h)$ has the following asymptotic behavior,
\begin{equation}\label{asym_horizon}
    R(h)\sim (h-h_+)^{\pm
    i\alpha},
\end{equation}
where $\alpha$ is a parameter to be fixed and $h_+$ and $h_-$ are determined by Eq.~\eqref{eq:horizon2} for rLQGBHs. 
Then we substitute Eq.~\eqref{asym_horizon} into Eq.~\eqref{radialh} and solve it.
Thus, we derive the specific form of $\alpha$,
\begin{equation}
    \alpha=\frac{a m-2M_{\rm B}\omega( h_+-3A_\lambda M_{\rm B})}{\epsilon_+ (h_+-h_-)}, \qquad \epsilon_+=\sqrt{1-\frac{8A_\lambda M_{\rm B}^2}{h_+^2}}.
\end{equation}
Near the spatial infinity, we perform an asymptotic expansion of Eq.~\eqref{radialh} and find that $R(h)$ takes the form,
\begin{equation}
    	R(h)\sim \frac{1}{h}\mathrm e^{\pm qh}h^{\pm M_{\rm B}(\mu^2-2\omega^2)/q},\qquad q=\sqrt{\mu^2-\omega^2}.\label{asym_infinity}
    \end{equation}
In Eq.~\eqref{asym_horizon} the plus and minus signs correspond to ingoing and outgoing waves, respectively. In contrast, the plus and minus signs correspond to outgoing and ingoing waves in Eq.~\eqref{asym_infinity}, respectively.
    
%The QNMs are eigenvalues of wave equations satisfying specific boundary conditions, where only ingoing waves exist near event horizons and only outgoing waves at the spatial infinity. Therefore, we only choose

QNMs are characterized as the eigenvalues of wave equations that adhere to stringent boundary conditions, stipulating the exclusive presence of ingoing waves in the vicinity of event horizons and solely outgoing waves at the spatial infinity. These boundary conditions profoundly influences the selection of suitable asymptotic behaviors for the wave function $R(h)$. Consequently, for an accurate representation of the wave dynamics near these critical regions, we adopt the following asymptotic expressions:
Near the outer event horizon, the radial wave function $R(h)$ asymptotically approaches:
    \begin{equation}\label{eq:as_horizon}
    	R(h)\sim (h-h_+)^{ i\alpha},
    \end{equation}
and near the spatial infinity, the asymptotic behavior takes the form: 
    \begin{equation}\label{eq:as_spinfty}
    	R(h)\sim \frac{1}{h}\mathrm e^{ qh}h^{M_{\rm B}(\mu^2-2\omega^2)/q}.
    \end{equation}
    
%Now we can determine the QNMs with the shooting method in the range of  $h_+<h<h_\infty$. In the first step, we choose a QNM frequency  as our initial value, with which we can determine the angular eigenvalue $A_{lm}$ by using the Leaver method~\cite{leaver1985analytic}. In the second step, considering the asymptotic behavior Eq.~\eqref{eq:as_horizon} near the outer horizon, we integrate Eq.~\eqref{radialh} from the outer event horizon to an intermediate point, where this point is usually chosen with the maximum value of the potential.  In the third step, considering the asymptotic behavior Eq.~\eqref{eq:as_spinfty} near the infinity, we integrate Eq.~\eqref{radialh} from the infinity to this intermediate point. In the final step, we require that the radial wave function $R(h)$ and its first derivative $R'(h)$ are continuous\footnotemark[1]\footnotetext[1]{Here ``continuous" means that the radial wave function $R(h)$ obtained in the second step equals that in the third step, and so does the first derivative of $R(h)$.} at the intermediate point and thus obtain a QNM frequency. Regarding this QNM frequency as the initial value for the next calculation, i.e. after an iterative process we at last get a stable QNM frequency.
Utilizing the shooting method, we can now accurately compute the QNMs in the range of $h_+<h<h_\infty$. Initially, we select a preliminary QNM frequency to serve as our initial guess. With this frequency we can deternime the angular eigenvalue $A_{lm}$ by the Leaver method~\cite{leaver1985analytic}.
Subsequently, in alignment with the asymptotic behavior delineated by Eq.~\eqref{eq:as_horizon} in the proximity of the outer event horizon, we  numerically integrate Eq.~\eqref{radialh}, commencing from the outer event horizon to a strategically chosen intermediate location. This point is typically selected to coincide with the peak of the potential function, optimizing the accuracy of our integration process.
In the ensuing phase, considering the asymptotic behavior presented in Eq.~\eqref{eq:as_spinfty} near the spatial infinity, we integrate Eq.~\eqref{radialh} from infinity back to the aforementioned intermediate point.
The last step is to ensure the continuity for both the radial wave function $R(h)$ and its first derivative $R'(h)$ at the intermediate junction. Achieving this continuity allows us to ascertain a QNM frequency. This newly determined QNM frequency is then employed as the initial value for the subsequent iteration. Through this iterative refinement, we ultimately converge upon a stable QNM frequency, thus fulfilling our objective of accurately determining the QNMs.

%The shooting method is very stable so that it can handle some situations in which the WKB method fails. However, it is not easy in the shooting method to give an initial value of QNMs and to make sure that it does not deviate from the expected value of QNMs too large. Otherwise, the shooting method does not work well. In practical applications, the first initial value will be fixed with the help of other numerical methods, and then the result from the previous iteration is regarded as the initial value for the next iteration.

The shooting method stands out for its exceptional stability, enabling it to handle the situations in which the Prony method and WKB method fail. However, the outcomes of the shooting method are highly dependent on the selection of initial values. In calculations, the initial values of QNMs are typically determined~\cite{leaver1985analytic} by the QNMs of Kerr black holes. With the above setup, we are able to compute QNMs through the shooting method.

%%%%%%%%%%%%%%%%%%%%%%%%%%%%%%%%%%%%%%%%%%%%%%%%%%%%%
\subsection{Numerical results}\label{sec:num-result}
%%%%%%%%%%%%%%%%%%%%%%%%%%%%%%%%%%%%%%%%%%%%%%%%%%%%%
 
%%%%%%%%%%%%%%%%%%%%%%%%%%%%%%%%%%%%%
\subsubsection{Comparison of accuracy among three methods}\label{sec:MPC}
%%%%%%%%%%%%%%%%%%%%%%%%%%%%%%%%%%%%%
%With the above three methods, we are able to obtain the QNMs of scalar field perturbations around rLQGBHs. Here we start by presenting the QNMs of massless scalar field perturbations with a varying $A_\lambda$ in Tab.~\ref{tab:QNM1}, where two modes with $l=1=m$ and $l=2=m$ are chosen. In the special case of $A_\lambda=0$, an rLQGBH reduces to a Kerr black hole, and our results are consistent with those computed by the Leaver method~\cite{konoplya2006stability}. 

Utilizing the three methodologies described, we successfully compute the QNMs for scalar field perturbations in the background of rLQGBHs. We commence our discussion by showcasing the QNMs of massless scalar field perturbations, which vary with the regularization parameter $A_\lambda$, as delineated in Tab.~\ref{tab:QNM1}. Specifically, we examine two modes, namely $l=1=m$ and $l=2=m$.
In the notable instance with $A_\lambda=0$, an rLQGBH simplifies to a Kerr black hole. It is noteworthy that our results align with those obtained~\cite{konoplya2006stability} via the Leaver method for Kerr black holes, thereby underscoring the accuracy and reliability of our analysis.

Since the Prony method is completely a numerical calculation of waveform simulations, its accuracy and reliability are the highest among the three methods, so we take its results as the standard to measure the accuracy of the results from the other two methods.
Here, we define the relative error between data $D$ and standard data $SD$ as follows:
\begin{equation}
     {\rm Err}=\left|\frac{D-SD}{D}\right|,
\end{equation}
and display the relative errors between the QNMs obtained by the Prony method and those by the other two methods in the brackets
of Tab.~\ref{tab:QNM1}, where the left of brackets shows the relative error of real parts, while the right that of imaginary parts. We can see that
the relative errors are almost all less than $2\%$, indicating that both the WKB method and shooting method exhibit high accuracy within the range of regularization parameter $A_\lambda$, $0\le A_\lambda \le 0.45$.
However, in the mode of $l=1=m$, the relative errors associated with the WKB method grow with an increase of regularization parameter $A_\lambda$, which implies that this method may not work well when other parameters, such as angular momenta, take a certain range.

In order to further study the scope of application of the WKB method and the impact of angular momentum $a$ on the QNMs of massless ({\em massive}) scalar field perturbations, we display in Tab.~\ref{tab:QNM2} the relationship between the QNMs and $a$, where $A_\lambda=0.1$ and $l=1=m$ are set. When $a\lesssim0.4$, the QNMs obtained from the three ({\em two}) methods are consistent. 
However, if $a\gtrsim0.4$, the discrepancy between the WKB method and the other two methods ({\em shooting method}) rapidly increases. 
The reason is that the series of $(a\omega)$ no longer converges in the WKB method  if $a\gtrsim0.4$, rendering the WKB method inapplicable. 

As shown in Tab. \ref{tab:QNM1}, the shooting method yields less relative errors than the 4th-order WKB, showing that the former exhibits higher precision than the latter.
Moreover, considering that the shooting method is much more efficient than the Prony method because the latter relies on complicated waveform simulations, we 
therefore prefer to apply the shooting method to the calculation of amplification factors in Sec.~\ref{sec:superradiance}.

\begin{table}[t]
\centering
\begin{tabular}{|l|c|c|c|}
\hline
\multicolumn{4}{|c|}{$l=1=m$}\\
\hline
$A_\lambda$ & Prony method&4th-order WKB&Shooting method\\ \hline
%精确值为0.301 045-0.097 547I
0&$0.3010-0.0975i $&$0.3011 - 0.0973 i \;(0.03\%, 0.20\%)$&$0.3010 - 0.0973 i \;(0.00\%, 0.20\%)$\\\hline
0.05&$ 0.3119-0.0998i$&$0.3119 - 0.0996 i \;(0.00\%, 0.20\%)$&$0.3118 - 0.0996 i \;(0.03\%, 0.20\%)$\\ \hline
0.10&$0.3242-0.1023i $&$0.3242 - 0.1022 i \;(0.00\%, 0.10\%)$&$0.3240 - 0.1023 i \;(0.06\%, 0.00\%)$\\\hline
0.15&$ 0.3382-0.1050i$&$0.3382 - 0.1049 i \;(0.00\%, 0.10\%)$&$0.3381 - 0.1053 i \;(0.03\%, 0.29\%)$\\\hline
0.20&$0.3544-0.1080i $&$0.3545 - 0.1077 i \;(0.03\%, 0.28\%)$&$0.3546 - 0.1082 i \;(0.06\%, 0.18\%)$\\\hline
0.25&$0.3735-0.1114i $&$0.3738 - 0.1108 i \;(0.08\%, 0.54\%)$&$0.3738 - 0.1112 i \;(0.08\%, 0.18\%)$\\\hline
0.30&$0.3964-0.1150i $&$0.3968 - 0.1140 i \;(0.10\%, 0.87\%)$&$0.3962 - 0.1147 i\;(0.05\%, 0.26\%)$\\\hline
0.35&$0.4245-0.1190i $&$0.4252 - 0.1174 i \;(0.16\%, 1.34\%)$&$0.4243 - 0.1193 i\;(0.05\%, 0.25\%)$\\\hline
0.40&$0.4604-0.1229i $&$0.4612 - 0.1207 i \;(0.17\%, 1.79\%)$&$0.4608 - 0.1227 i\;(0.09\%, 0.16\%)$\\\hline
0.45&$0.5082-0.1256i $&$0.5093 - 0.1232 i \;(0.22\%, 1.91\%)$&$0.5079 - 0.1258 i\;(0.06\%, 0.16\%)$\\\hline
\end{tabular}
\begin{tabular}{|l|c|c|c|}
\hline
\multicolumn{4}{|c|}{$l=2=m$}\\
\hline
$A_\lambda$ & Prony method&4th-order WKB&Shooting method\\ \hline
0   &$0.4995-0.0967i $&$0.4995 - 0.0966  i\;(0.00\%, 0.10\%)$&$0.4994 - 0.0967 i\;(0.02\%, 0.00\%)$\\\hline
0.05&$0.5179-0.0989i $&$0.5179 - 0.0989 i \;(0.00\%, 0.00\%)$&$0.5180 - 0.0990 i\;(0.02\%, 0.10\%)$\\ \hline
0.10&$0.5388-0.1015i $&$0.5388 - 0.1014  i\;(0.00\%, 0.10\%)$&$0.5389 - 0.1014 i\;(0.02\%, 0.10\%)$\\\hline
0.15&$0.5627-0.1042i $&$0.5627 - 0.1042  i\;(0.00\%, 0.00\%)$&$0.5626 - 0.1042 i\;(0.02\%, 0.00\%)$\\\hline
0.20&$0.5904-0.1073i $&$0.5904 - 0.1073 i\;(0.00\%, 0.00\%)$&$0.5903 - 0.1074 i\;(0.02\%, 0.09\%)$\\\hline
0.25&$0.6231-0.1108i $&$0.6231 - 0.1107 i\;(0.00\%, 0.09\%)$&$0.6232 - 0.1107 i\;(0.02\%, 0.09\%)$\\\hline
0.30&$0.6626-0.1146i $&$0.6626 - 0.1145 i\;(0.00\%, 0.09\%)$&$0.6624 - 0.1146 i\;(0.03\%, 0.00\%)$\\\hline
0.35&$0.7114-0.1187i$&$0.7115 - 0.1185 i\;(0.01\%, 0.17\%)$&$0.7116 - 0.1187 i\;(0.03\%, 0.00\%)$\\\hline
0.40&$0.7743-0.1229i$&$0.7743 - 0.1225 i \;(0.00\%, 0.32\%)$&$0.7743 - 0.1229 i\;(0.00\%, 0.00\%)$\\\hline
0.45&$0.8588-0.1255i$&$0.8587 - 0.1250 i\;(0.01\%, 0.40\%)$&$0.8588 - 0.1257 i\;(0.00\%, 0.16\%)$\\\hline
\end{tabular}
\caption{QNMs for the modes of $l=1=m$ and $l=2=m$, respectively, where  $a=0.1$, $M_{\rm B}=1$, and $\mu=0$ are set.}
\label{tab:QNM1}
\end{table}

\begin{table}
\centering
\fontsize{11pt}{12pt}\selectfont
\begin{tabular}{|l|c|c|c|c|c|}
\hline
&\multicolumn{3}{|c|}{$\mu=0$}
&\multicolumn{2}{|c|}{$\mu=0.1$}
\\
\hline
$a$&Prony method &4th-order WKB&Shooting method&4th-order WKB&Shooting method\\ \hline
0&$ 0.3145-0.1024i$&$0.3146 - 0.1022 i $&$0.3144 - 0.1022 i$&$0.3187 - 0.0998 i$&$0.3186 - 0.0996 i$\\\hline
0.1&$0.3242-0.1023i$&$ 0.3242 - 0.1022 i$&$0.3240 - 0.1023 i$&$0.3281 - 0.0999 i$&$0.3278 - 0.0999 i$\\ \hline
0.2&$0.3351-0.1020i$&$0.3351 - 0.1019 i $&$0.3349 - 0.1022 i$&$0.3388 - 0.0998 i$&$0.3385 - 0.1000 i$\\\hline
0.3&$0.3474-0.1014i$&$ 0.3474 - 0.1012 i$&$ 0.3475-0.1016i$&$0.3509 - 0.0993 i $&$0.3509 - 0.0998 i$\\\hline
0.4&$0.3617-0.1004i$&$0.3617 - 0.0995 i $&$0.3619 - 0.1004 i$&$0.3649 - 0.0978 i$&$0.3651 - 0.0989 i$\\\hline
0.5&$0.3785-0.0987i$& $0.3782 - 0.0939 i$&$0.3785 - 0.0986 i$&$0.3808 - 0.0923 i$&$0.3816 - 0.0972 i$\\\hline
0.6&$0.3987-0.0962i$&$ 0.3909 - 0.0761 i$&$0.3986 - 0.0961 i$&$0.3924 - 0.0752 i$&$0.4013 - 0.0949 i$\\\hline
0.7&$0.4242-0.0921i$&$0.3849 - 0.0530 i$&$0.4242 - 0.0921 i$&$0.3860 - 0.0530 i$&$0.4265 - 0.0912 i$\\\hline
0.8&$0.4585-0.0850i$&$0.3693 - 0.0380 i$&$0.4585 - 0.0849 i$&$0.3697 - 0.0381 i$&$0.4605 - 0.0843 i$\\\hline
0.9&$0.5116-0.0706i$&$0.3535 - 0.0292 i$&$0.5116-0.0706 i$&$0.3529 - 0.0292 i$&$0.5129 - 0.0703 i$\\\hline
\end{tabular}
\caption{QNMs for the mode of $l=1=m$ with $A_\lambda=0.1$, $a=0.1$, $M_{\rm B}=1$, $\mu=0$ or $\mu=0.1$. Note that the Prony method is unsuitable for massive scalar field perturbations, so no corresponding data are shown.}
\label{tab:QNM2}
\end{table}

%%%%%%%%%%%%%%%%%%%%%%%%%%%%%%%%%%%%%%%%%%%%%%%%%%%%%%%%%%%%%%%%%%%%%%%%%
\subsubsection{Data of quasi-normal modes}
%%%%%%%%%%%%%%%%%%%%%%%%%%%%%%%%%%%%%%%%%%%%%%%%%%%%%%%%%%%%%%%%%%%%%%%%%
\begin{figure}[t]
\centering
    	\subfigure[]{
    		\begin{minipage}[t]{0.4\linewidth}
    			\centering\label{fig:QNMR-A-a-2D}
    			\includegraphics[width=1\linewidth]{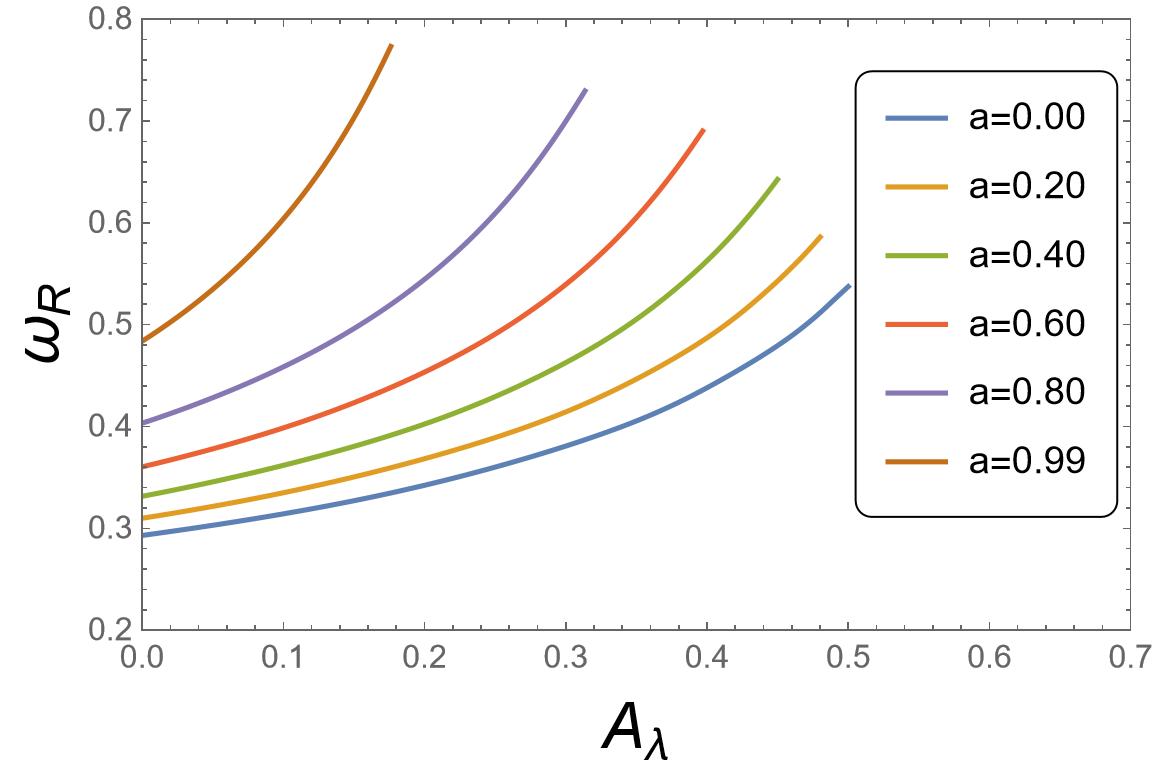}
    		\end{minipage}
    	}
\subfigure[]{
\begin{minipage}[t]{0.4\linewidth}
\centering\label{fig:QNMR-A-a-3D}
\includegraphics[width=1\linewidth]{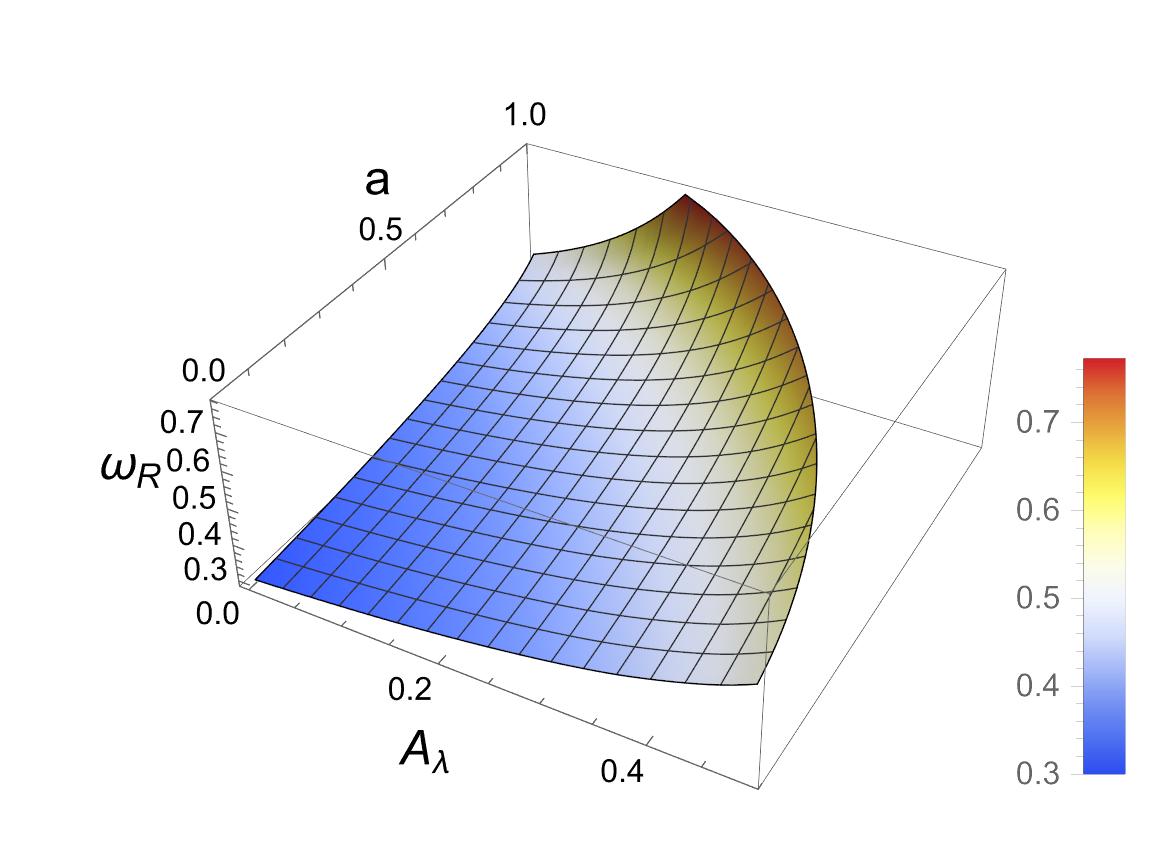}
\end{minipage}
    	}
\subfigure[]{
\begin{minipage}[t]{0.4\linewidth}
\centering\label{fig:QNMI-A-a-2D}
\includegraphics[width=1\linewidth]{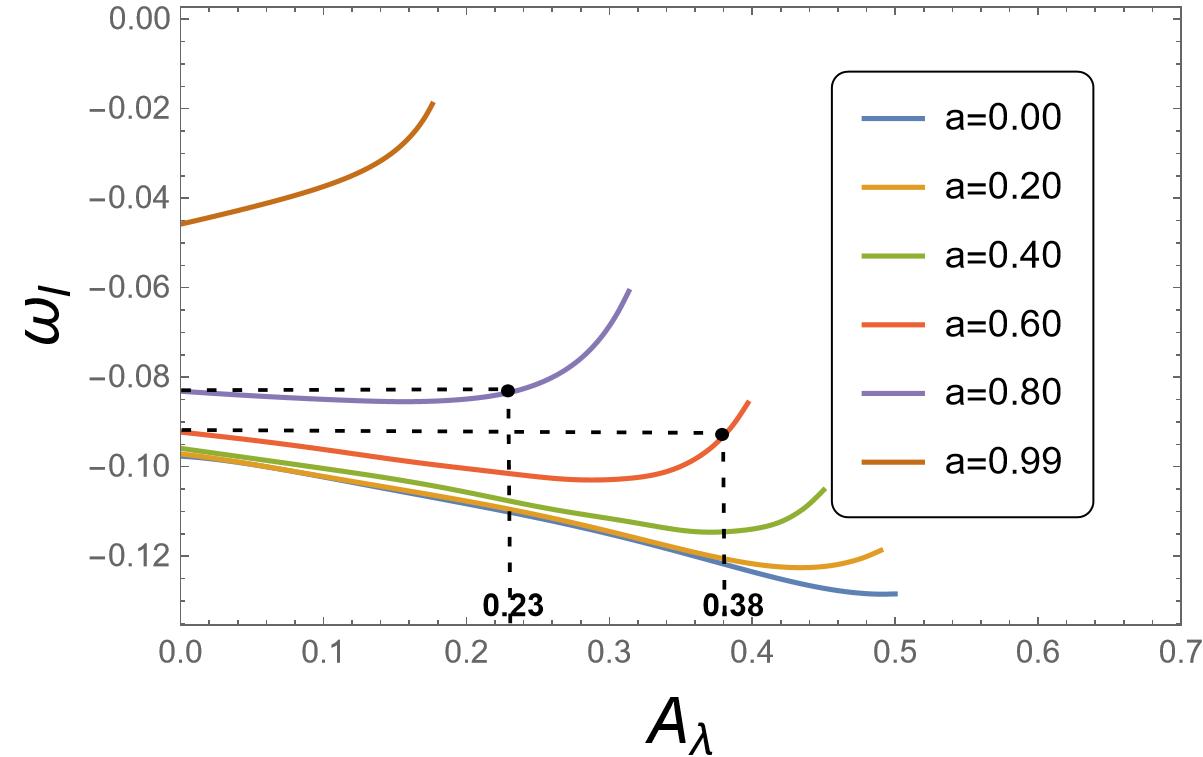}
\end{minipage}
    	}
\subfigure[]{
\begin{minipage}[t]{0.4\linewidth}
\centering\label{fig:QNMI-A-a-3D}
\includegraphics[width=1\linewidth]{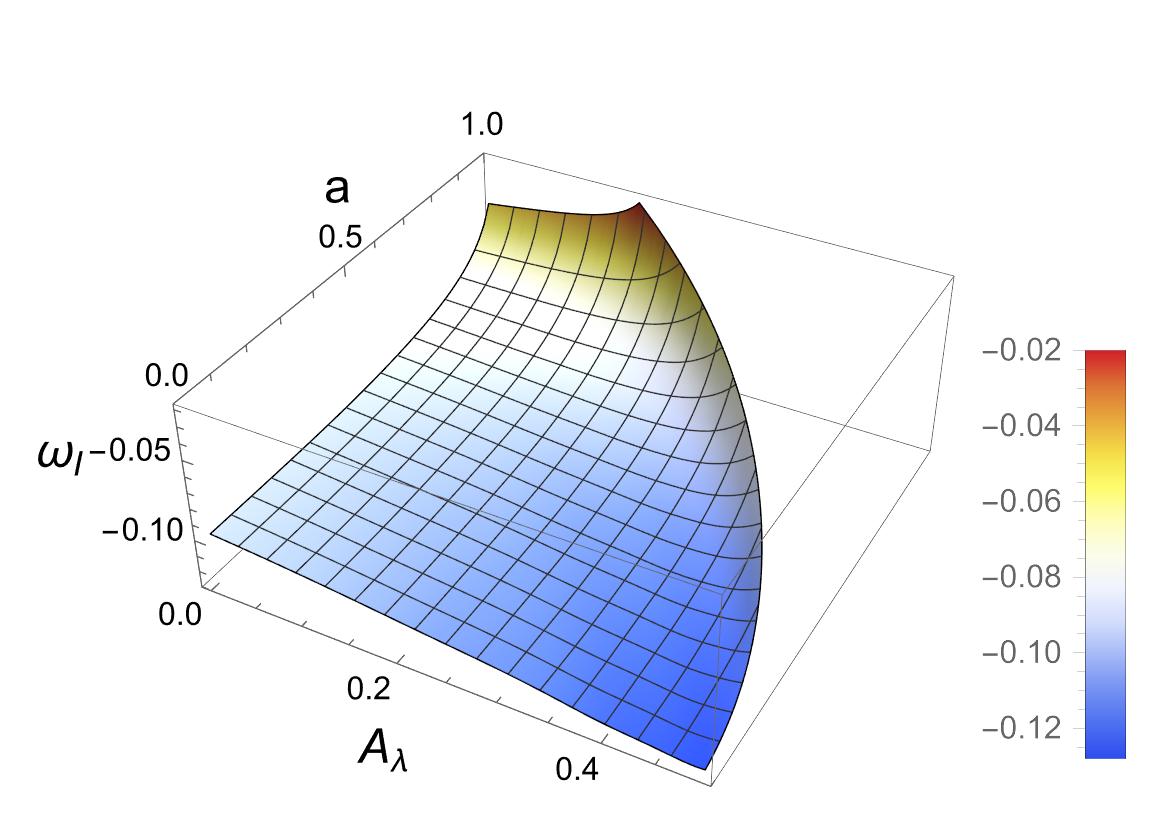}
\end{minipage}
    	}
\caption{QNM frequencies $\omega$ as a function of regularization parameter $A_\lambda$ with a varying angular momentum, $a=0$, $0.2$, $0.4$, $0.6$, $0.8$, and $0.99$, where $M_{\rm B}=1$ and $\mu=0$ are set, and the mode of  $l=1=m$ is chosen. In diagram (c), $|\omega_{\rm I}|$ of rLQGBHs is smaller than that of Kerr black holes ($A_\lambda=0$) when $A_\lambda$ is greater than 0.38 for the case of $a=0.60$, and the same phenomenon happens when $A_\lambda$ is greater than 0.23 for the case of $a=0.80$.}
\label{fig:QNM-A-a}
    \end{figure}

Our analyses reveal that the regularization parameter $A_\lambda$ has a significant impact on the QNMs of massless scalar field perturbations
in Tab.~\ref{tab:QNM1}, where we present the data on how the QNMs change with respect to the regularization parameter $A_\lambda$ for both $l=1=m$ and $l=2=m$ modes when $a=0.1$ and $M_{\rm B}=1$ are set.
In the two modes, the real parts $\omega_{\rm R}$ grow with an increase of the regularization parameter $A_\lambda$. 
This indicates that an increase of $A_\lambda$ amplifies the oscillation frequencies of scalar fields. For example, 
in the mode of $l=1=m$, the real part corresponding to $A_\lambda=0.45$ is 1.69 times larger than that corresponding to $A_\lambda=0$.
Furthermore, as $A_\lambda$ increases, the absolute value of imaginary parts $|\omega_{\rm I}|$ also increases, indicating a faster dissipation of waveforms.
For instance, in the mode of $l=1=m$, $|\omega_{\rm I}|$ corresponding to $A_\lambda=0.45$ is 1.29 times larger than that corresponding to $A_\lambda=0$.
These results highlight the significance of regularization parameter $A_\lambda$ in influencing the behaviors of QNMs of massless scalar field perturbations. 
%which shows that the slope between adjacent peaks on the damping oscillation waveform under $A_\lambda=0.45$ is 1.29 times that under $A_\lambda=0$.
% which shows that the time interval between adjacent peaks on the damping oscillation waveform under $A_\lambda=0.45$ is only $59\%$ of that under $A_\lambda=0$.
Now we present the influence of $A_\lambda$ on the QNMs of massless scalar field perturbations under different values of $a$ in Fig.~\ref{fig:QNM-A-a}, where the maximum value of $A_\lambda$ is determined by Eq.~\eqref{max}.
Here we take the mode of  $l=1=m$ as an example and set $M_{\rm B}=1$.
As shown in Figs.~\ref{fig:QNMR-A-a-2D} and \ref{fig:QNMR-A-a-3D}, the real parts of QNMs always increase with an increase of $A_\lambda$ under a fixed $a$, where
the largest real part is $\omega_{\rm R}=0.7828$, located at $a=1$ and $A_\lambda=0.125$, and it corresponds to the extreme configuration of rLQGBHs with the greatest angular momentum, $a_{\rm Max}=1$.
Moreover, as shown in Figs.~\ref{fig:QNMI-A-a-2D} and \ref{fig:QNMI-A-a-3D}, $A_\lambda$ has diverse effects on $|\omega_{\rm I}|$ under different values of $a$, which can be divided into three categories:
\begin{itemize}
    \item When $0<a\leq0.48$, $|\omega_{\rm I}|$ increases at first and then decreases when $A_\lambda$ grows. 
    Within the entire range of $A_\lambda$, $|\omega_{\rm I}|$ is always larger than that corresponding to $A_\lambda=0$. 
    From a physical perspective, a non-vanishing $A_\lambda$ gives rise to a faster decay of massless scalar fields than the case of  vanishing $A_\lambda$, thereby making the spacetime more unstable.
    Note that an rLQGBH reaches its  extreme configuration when $A_\lambda$ takes its maximum value, see Eq.~\eqref{max}, and it reduces to a Kerr black hole when $A_\lambda$ equals zero. 
    When $a$ increases, the gap between $|\omega_{\rm I}|$ of an extreme rLQGBH and that of a Kerr black hole is getting smaller and smaller, and finally it disappears when $a=0.48$.
    \item When $0.48<a\leq0.82$, $|\omega_{\rm I}|$ also initially increases and then  decreases when $A_\lambda$ grows. 
    The inflection depends on the critical value of $A_\lambda$ that is associated with $a$. When $A_\lambda$ exceeds such a value, $|\omega_{\rm I}|$ is smaller than that of the Kerr case.
    In Fig.~\ref{fig:QNMI-A-a-2D}, we show that the critical value equals $0.38$ if $a=0.60$, and equals $0.23$ if $a=0.80$, respectively. 
    Therefore, $A_\lambda$ plays the role in promoting the stability of spacetime when $A_\lambda$ is larger than this value,
    but it plays the role in diminishing the stability of spacetime when $A_\lambda$ is smaller than this value.
    Additionally, this critical value becomes small when $a$ increases, and it reaches zero when $a=0.82$.
    
    \item When $0.82<a\leq1.00$, $|\omega_{\rm I}|$ decreases monotonically when $A_\lambda$ grows, implying that
    $A_\lambda$ plays the role in promoting the stability of spacetime.
     When $a=1$ and $A_\lambda=0.125$, $|\omega_{\rm I}|$ reaches the minimum value, $|\omega_{\rm I}|_{\rm Min}=0.0202$.
\end{itemize}

\begin{figure}[t]
\centering
    	\subfigure[]{
    		\begin{minipage}[t]{0.4\linewidth}
    			\centering\label{fig:QNMR-A-a-2-2D}
    			\includegraphics[width=1\linewidth]{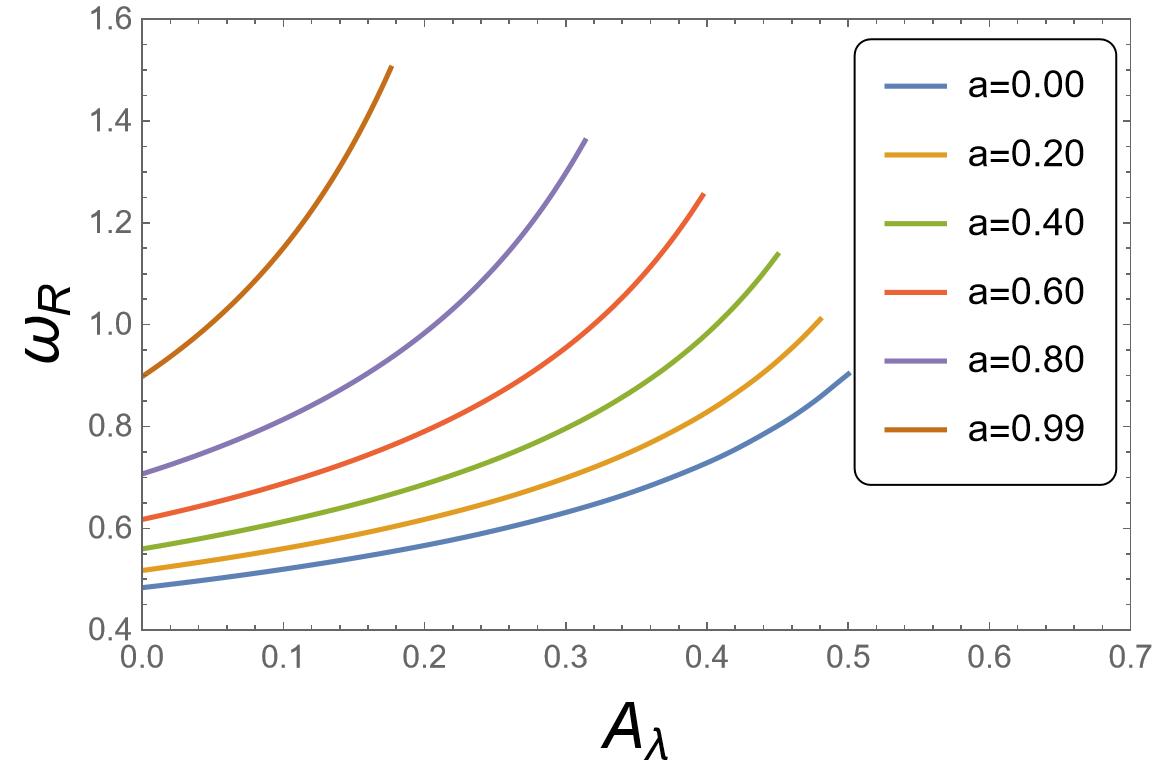}
    		\end{minipage}
    	}
\subfigure[]{
\begin{minipage}[t]{0.4\linewidth}
\centering\label{fig:QNMR-A-a-2-3D}
\includegraphics[width=1\linewidth]{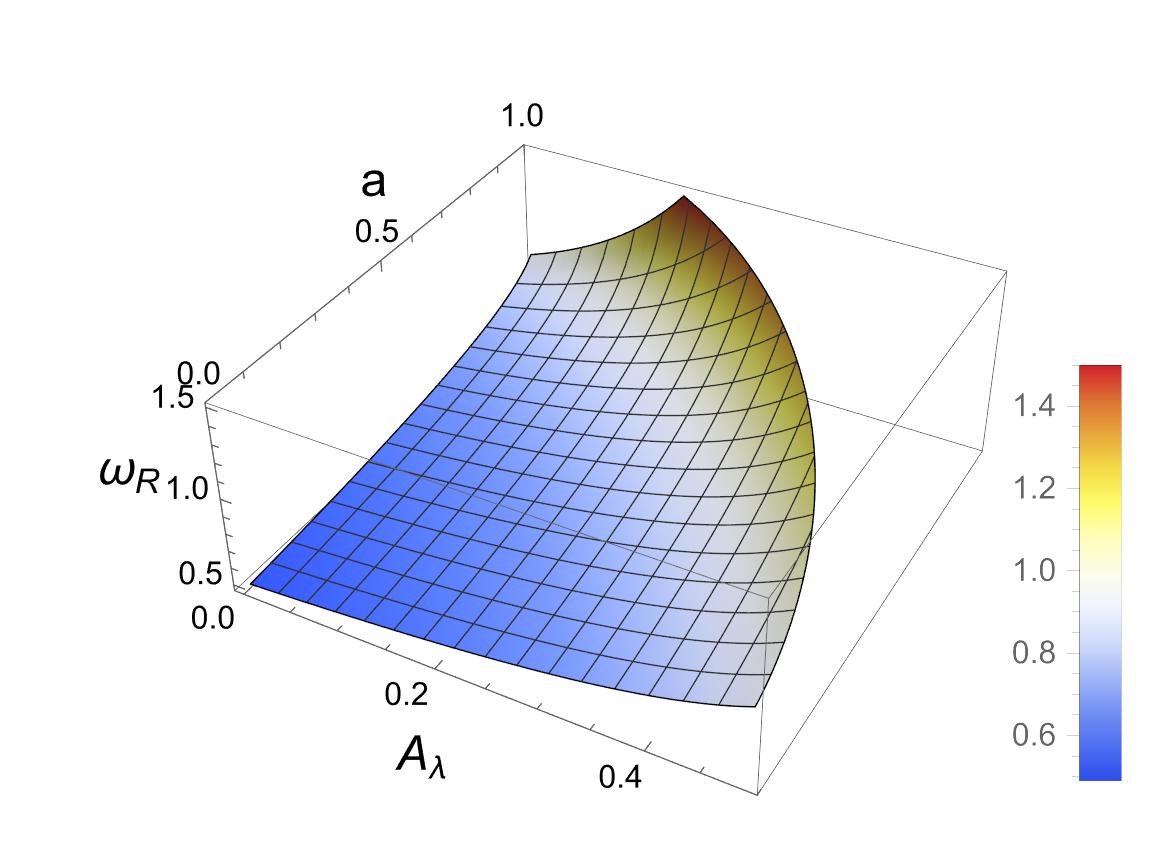}
\end{minipage}
    	}
\subfigure[]{
\begin{minipage}[t]{0.4\linewidth}
\centering\label{fig:QNMI-A-a-2-2D}
\includegraphics[width=1\linewidth]{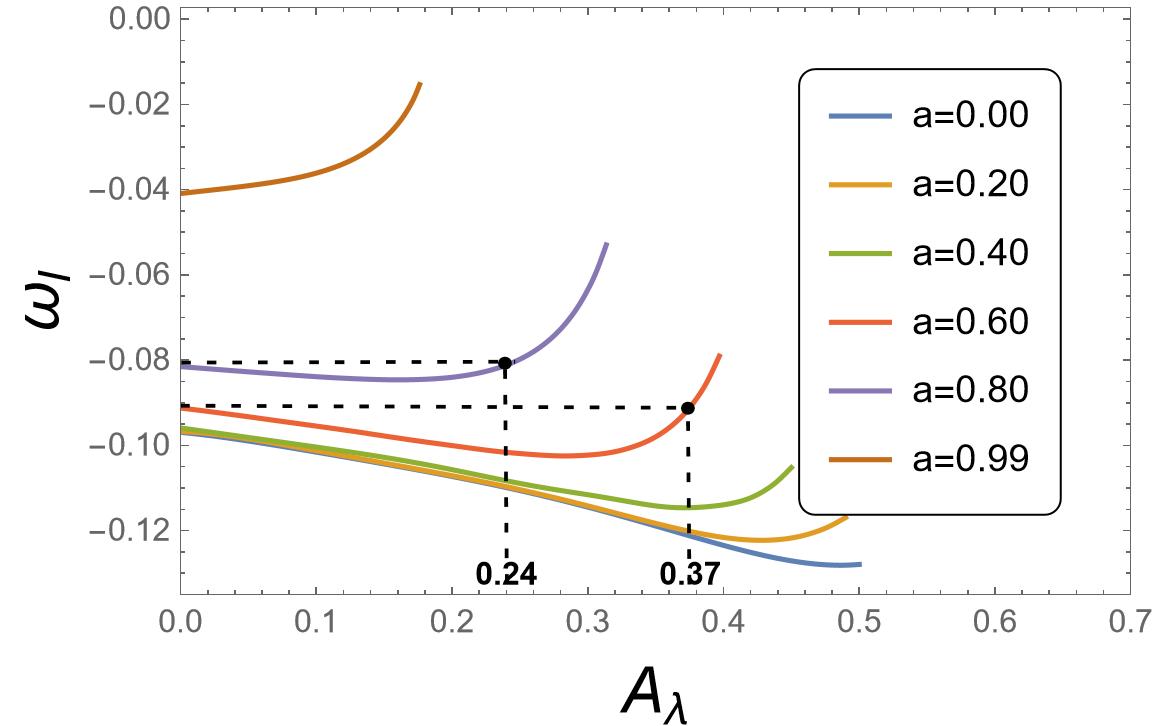}
\end{minipage}
    	}
\subfigure[]{
\begin{minipage}[t]{0.4\linewidth}
\centering\label{fig:QNMI-A-a-2-3D}
\includegraphics[width=1\linewidth]{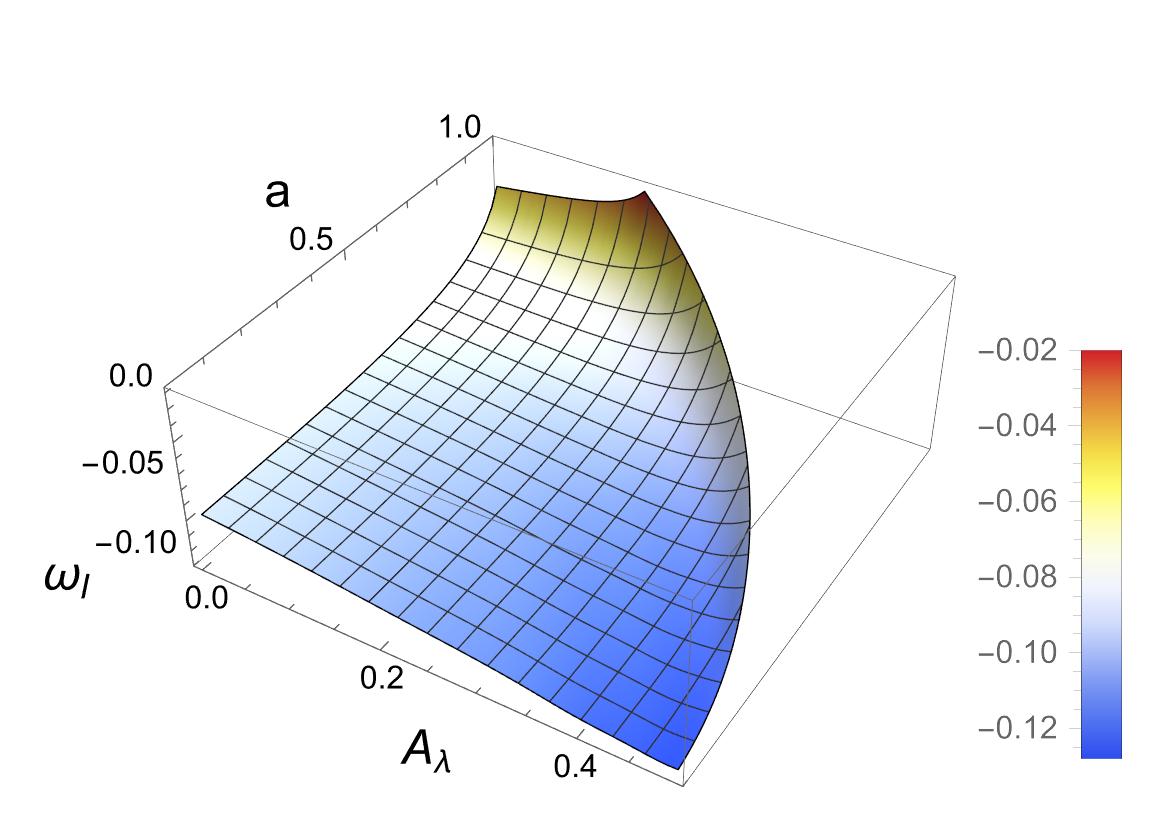}
\end{minipage}
    	}
\caption{QNM frequencies $\omega$ as a function of regularization parameter $A_\lambda$ with a varying angular momentum, $a=0$, $0.2$, $0.4$, $0.6$, $0.8$, and $0.99$, where $M_{\rm B}=1$ and $\mu=0$ are set, and the mode of $l=2=m$ is chosen. In diagram (c), $|\omega_{\rm I}|$ of rLQGBHs is smaller than that of Kerr black holes ($A_\lambda=0$) when $A_\lambda$ is greater than 0.37 for the case of $a=0.60$, and the same phenomenon happens when $A_\lambda$ is greater than 0.24 for the case of $a=0.80$.}
\label{fig:QNM-A-a-2}
    \end{figure}

In addition, we present the influence of $A_\lambda$ on the QNMs of massless scalar field perturbations under different values of $a$ for the mode of $l=2=m$ in Fig.~\ref{fig:QNM-A-a-2}, where the maximum value of $A_\lambda$ is also determined by Eq.~\eqref{max}.
In general, the influence of $A_\lambda$ on QNMs in the mode of $l=2=m$ is similar to that in the mode of $l=1=m$, i.e., the former differs from the latter just in  numerical differences.
For the real parts of QNMs, the former is significantly larger than the latter.
But for the absolute value of imaginary parts, the difference between the two modes is very small.
Similarly, for the mode of $l=2=m$, the influence of $A_\lambda$ on $|\omega_{\rm I}|$ can also be divided into three categories for a varying $a$: $0<a\leq0.49$, $0.49<a\leq0.84$, and $0.84<a\leq1.00$.
From Fig.~\ref{fig:QNM-A-a-2} it is clear that the relations between $\omega_{\rm R}$ ($|\omega_{\rm I}|$) and $A_\lambda$ are quite similar in the two modes.

\begin{figure}[t]
\centering
\subfigure[]{\begin{minipage}[t]{0.4\linewidth}
\centering
\includegraphics[width=1\linewidth]{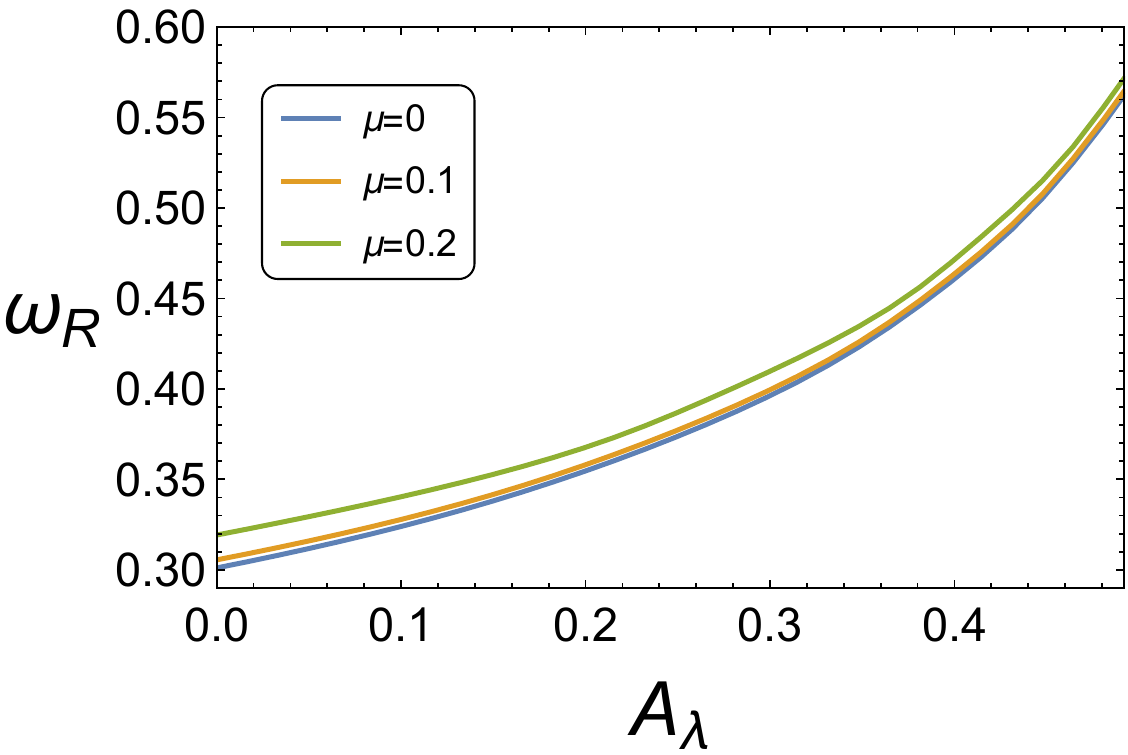}
\end{minipage}
    	}
\subfigure[]{
\begin{minipage}[t]{0.4\linewidth}
\centering
\includegraphics[width=1\linewidth]{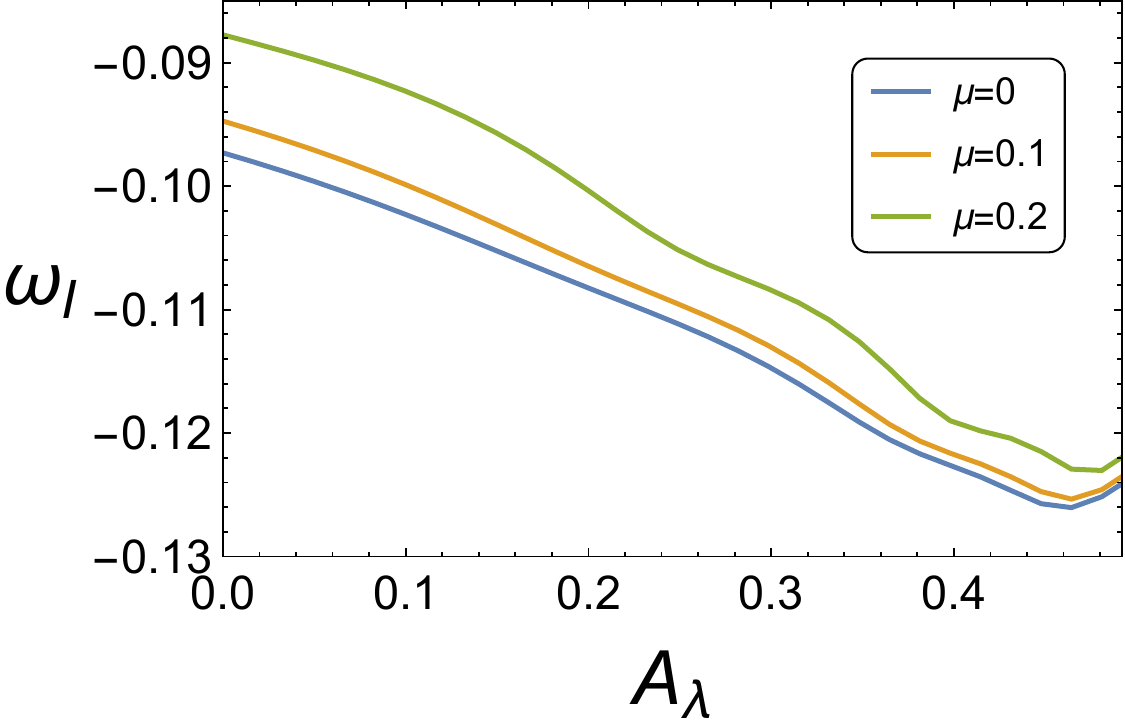}
\end{minipage}
    	}
\subfigure[]{
\begin{minipage}[t]{0.4\linewidth}
\centering
\includegraphics[width=1\linewidth]{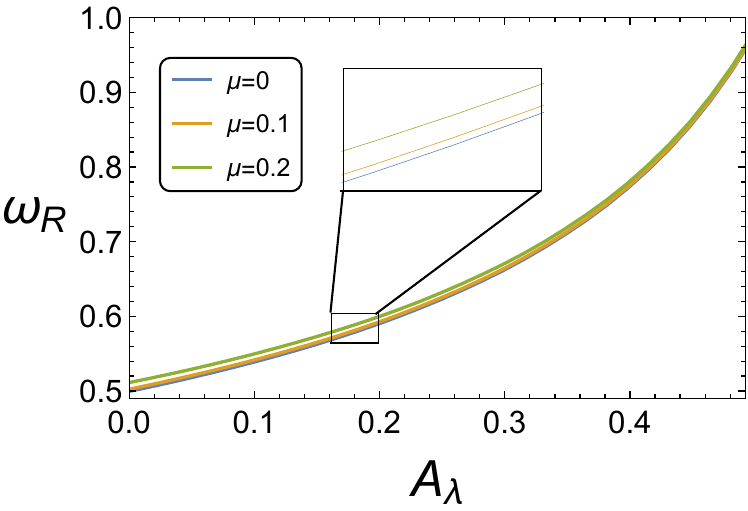}
\end{minipage}
    	}
\subfigure[]{
\begin{minipage}[t]{0.4\linewidth}
\centering
\includegraphics[width=1\linewidth]{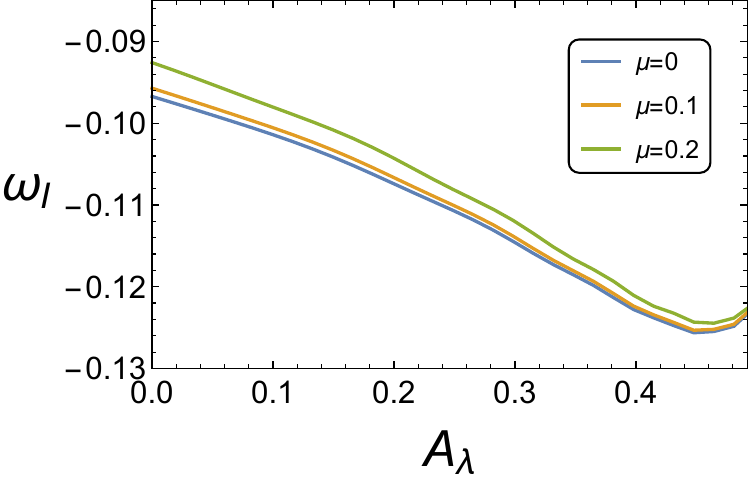}
\end{minipage}
    	}
\caption{QNM frequencies $\omega$ as a function of regularization parameter $A_\lambda$ with different masses of scalar fields, $\mu=0$, $0.1$, and $0.2$, where $a=0.1$, $M_{\rm B}=1$, and $n=0$ are set. The upper two diagrams correspond to the mode of $l=1=m$, and the lower two diagrams the mode of $l=2=m$.}
\label{mu_omega}
\end{figure}

In Fig.~\ref{mu_omega}, we demonstrate the relationship between the QNMs and regularization parameter $A_\lambda$ under a varying scalar field mass $\mu$. 
It can be observed for a fixed $A_\lambda$ that the presence of $\mu$ leads to an increase of the real parts of $\omega$ but a decrease of the absolute value of imaginary parts. 
This implies that the massive scalar field perturbations around rLQGBHs oscillate  faster but decay more slowly compared to the massless case, and that the mass $\mu$ plays the role in promoting the stability of spacetime.
When $A_\lambda=0$, i.e., in Kerr black holes $\mu$ has the greatest influence on the QNMs. 
When $A_\lambda$ gradually increases, the influence of $\mu$ gradually decreases. 
Therefore, a non-vanishing $A_\lambda$ makes $\mu$ have a weaker effect on promoting the spacetime stability than a vanishing $A_\lambda$  does.

At the end of this section, we make a comment that a pure imaginary frequency means a mode that has no oscillation but just damping. In our study for rLQGBHs, we do not find any pure imaginary frequency in the damping oscillation stage. According to the results shown in Tabs.~\ref{tab:QNM1} and \ref{tab:QNM2} together with Figs.~\ref{fig:QNM-A-a}, \ref{fig:QNM-A-a-2}, and \ref{mu_omega}, it is evident that the real parts increase when $A_\lambda$ and $a$ increase, and that they can never go to zero even if $A_\lambda$ and $a$ vanish. Specifically, for $n=0$ modes with $l=1=m$ and $l=2=m$, we can see that the real parts are non-zero for Schwarzschild black holes with vanishing $A_\lambda$ and $a$, and that the real parts of such modes for rLQGBHs become large when $A_\lambda$ and $a$ are increasing.
   
%%%%%%%%%%%%%%%%%%%%%%%%%%%%%%%%%%%%%%%%%%%%%%%%%%%%%%%%%%%%%%%%%%%%%%%%    
 \section{Superradiance}\label{sec:superradiance}
 %%%%%%%%%%%%%%%%%%%%%%%%%%%%%%%%%%%%%%%%%%%%%%%%%%%%%%%%%%%%%%%%%%%%%%%

%Penrose proposed the idea of extracting energy from black holes based on the presence of negative energy orbits within the ergosphere of rotating black holes.

The idea of extracting energy from rotating black holes was initially proposed by Penrose~\cite{penrose1971extraction}, which is based on the phenomenon of negative energy orbits within the ergosphere of black holes, now known as the Penrose process. However, the conditions to realize the Penrose process are so stringent that the process can hardly be observed in astrophysical phenomena. Inspired by the Penrose process, several more realistic mechanisms for energy extraction  were subsequently suggested, including the superradiance and Blandford-Znajek mechanism~\cite{komissarov2008blandford}, and so on. 
Among these mechanisms, the scalar superradiance recently attracts lots of attention. Unlike the Penrose process, the scalar superradiance involves ultralight scalar fields, where  such scalar fields are promising candidates of dark matter. The mechanism of superradiance exploits the interaction of scalar fields with rotating black holes to amplify the fields and extract energy, providing a fascinating sample of how quantum field theory concepts might be observable in astrophysical contexts. Owing to such a close relationship between the Penrose process and superradiance, we briefly introduce this process in the next subsection.

\subsection{The Penrose process}
The Penrose process involves particle dynamics near a black hole, and it is closely related to the phenomenon of superradiance.
Here we qualitatively infer the effects of the regularization parameter by analyzing the impact of this parameter on the Penrose process.
For a particle with mass $\mu_0$ that decays into two particles each with mass $\mu_{f}$, the efficiency of energy extraction can be expressed~\cite{chandrasekhar1998mathematical,Brito:2015oca} as follows:
\begin{equation}
\eta = \frac 12 \left[\sqrt{\left(1+g_{tt}\right)\left(1-\frac{4\mu_{f}^2}{\mu_0^2}\right)}-1 \right],\label{Peneff}
\end{equation}
where $g_{tt}$ means the $(t,t)$ component of the rLQGBH metric, see Eqs.~\eqref{1rotating_LQGBH_metric}, \eqref{eq:rho}, \eqref{areal}, \eqref{Mb}, and \eqref{eq:x}. The efficiency reaches its maximum when $g_{tt}$ takes its maximum value and  the two particles are photons with vanishing mass, $\mu_{f}=0$.

The metric of rLQGBHs differs from that of Kerr black holes because it incorporates an additional regularization parameter $A_\lambda$. As shown in Fig. \ref{fig:rLQG_parameter} which displays the parameter space of rLQGBHs, the range of $A_\lambda$ is constrained by the angular momentum $a$. Within the permissible parameter space, the maximal efficiency reads\footnote[2]{Substituting $\mu_{f}=0$ into Eq.~\eqref{Peneff} and scanning the parameter space $(A_\lambda, a)$ for searching the maximum value of $g_{tt}$, we thus reach the maximal efficiency. For the details of the parameter space, see Sec.~\ref{sec: Rotating_metric}.}
\begin{equation}
    \eta_{\mathrm{max}} = \frac 12 \left[\sqrt{\frac{2M_{\mathrm B} h_0-6A_{\lambda 0}M_{\mathrm B}^2}{ h_0^2-6A_{\lambda 0} M_{\mathrm B}^2}}-1 \right],
\end{equation}
where $h_0=M_{\mathrm B}$ and $A_{\lambda 0}=1/8$. As a result, we obtain $\eta_{\mathrm{max}} \approx 0.618$, which is significantly higher than the efficiency of $\eta_{\mathrm{max}} \approx 0.207$ for extreme Kerr black holes.

Through the above simple analysis to the Penrose process, we find that the introduction of  the regularization parameter results in a theoretically higher maximum efficiency of extractable energy from rLQGBHs compared to the Kerr scenario. Therefore, we can conclude that the energy extraction is relatively more economic from rLQGBHs than from Kerr black holes~\cite{chandrasekhar1998mathematical}.

\subsection{Numerical method for superradiance} 
When a free scalar field is incident on a black hole, the incident wave will be decomposed into two parts, a reflected wave and a transmitted wave, due to the scattering effect of potential barriers near the black hole.
The boundary conditions of the scattering process include pure ingoing wave at the event horizon and both ingoing wave and outgoing wave at the spatial infinity, where the ingoing wave at the event horizon  corresponds to the transmitted wave, while the ingoing wave and outgoing wave at the spatial infinity correspond to the incident wave and reflected wave, respectively.
The energy of reflected waves is greater than the energy of incident waves when the frequency $\omega$ of incident scalar fields satisfies~\cite{Yang:2022yvq} the following conditions:
\begin{equation}
     m\Omega_{\rm H}>\omega>\mu, \label{instability}
\end{equation}
where 
\begin{equation}\label{eq:angular-vec}
    \Omega_{\rm H}=\frac{a}{h_+^2-6A_\lambda M_{\rm B}^2+a^2}
\end{equation}
is the angular velocity of scalar particles at the outer event horizon $h_+$ determined by Eq.~\eqref{eq:horizon2} for rLQGBHs.
It is worth noting that the frequency $\omega$ of scalar fields is always real owing to the special boundary conditions of scattering processes.
This phenomenon of energy amplification is called~\cite{Brito:2015oca} superradiance.
Next we study how the energy amplification factor is affected by the regularization parameter $A_\lambda$.
 
%When a wave with a specific frequency and angular momentum approaches a rotating black hole, it can undergo a scattering process called superradiant scattering. In this process, the wave can extract energy from the black hole's rotational energy and be amplified, resulting in an exponential growth in amplitude. This is superradiance. It occurs due to the existence of a ergoregion near the black hole's event horizon, which allows for energy extraction.

%%%%%%%%%%%%%%%%%%%%%%%%%%%%%%%%%%%%%%%%%%%%%%%%%%%%%%%%%%%%%%%%
 
%%%%%%%%%%%%%%%%%%%%%%%%%%%%%%%%%%%%%%%%%%%%%%%%%%%%%%%%%%%%%%%%
In terms of the boundary conditions  mentioned above, the radial equation of motion  Eq.~\eqref{radialh} takes the asymptotic solution at the spatial infinity as follows:
\begin{equation}
   	R(h)\sim \mathscr{I} \frac{1}{h}\mathrm e^{- qh}h^{ -M_{\rm B}(\mu^2-2\omega^2)/q}+ \mathscr{R}\frac{1}{h} \mathrm e^{ qh}h^{ M_{\rm B}(\mu^2-2\omega^2)/q},\label{super_infinity}
   \end{equation}
where $\mathscr{I}$ and $\mathscr{R}$ stand for the incident and  reflection amplitudes, respectively. Moreover,
the asymptotic wave function near the outer horizon $h_+$ reads 
   \begin{equation}
   	R(h)\sim \mathscr{T}(h-h_+)^{i\alpha},
   \end{equation}
where $\mathscr{T}$ is the transmission amplitude. 
Then the amplification factor is given~\cite{Brito:2015oca} by 
\begin{equation}\label{eq:Z}
   	Z_{lm}=\frac{\mathrm dE_{\rm out}}{\mathrm dE_{\rm in}}=\left| \frac{\mathscr R}{\mathscr I}\right| ^2-1.
\end{equation} 
Following the shooting method outlined in Sec.~\ref{shooting}, we establish the relationship among $\mathscr{I}$, $\mathscr{R}$, and $\mathscr{T}$, and thus give the amplification factor of superradiance. 
%%%%%%%%%%%%%%%%%%%%%%%%%%%%%%%%%%%%%%%%%%%%%%%%%%%%%%%%%%%%%%%%%%
\subsection{Results}
%%%%%%%%%%%%%%%%%%%%%%%%%%%%%%%%%%%%%%%%%%%%%%%%%%%%%%%%%%%%%%%%%%

\begin{figure}[htbp]
   	\centering
   	\subfigure[]{
   		\begin{minipage}[t]{0.4\linewidth}
   			\centering \label{fig:rss-a}
   			\includegraphics[width=1\linewidth]{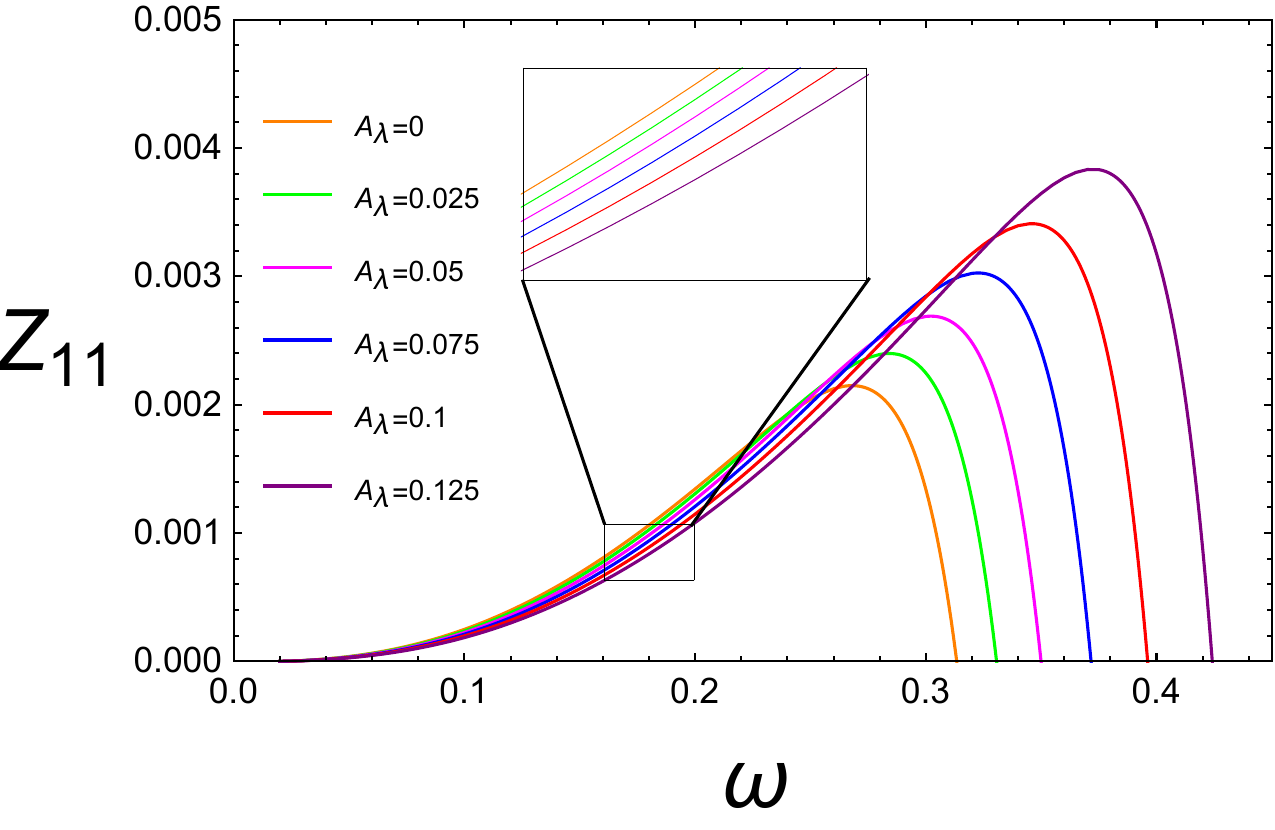}
   		\end{minipage}
   	}
   	\subfigure[]{
   		\begin{minipage}[t]{0.4\linewidth}
   			\centering \label{fig:rss-b}
   			\includegraphics[width=1\linewidth]{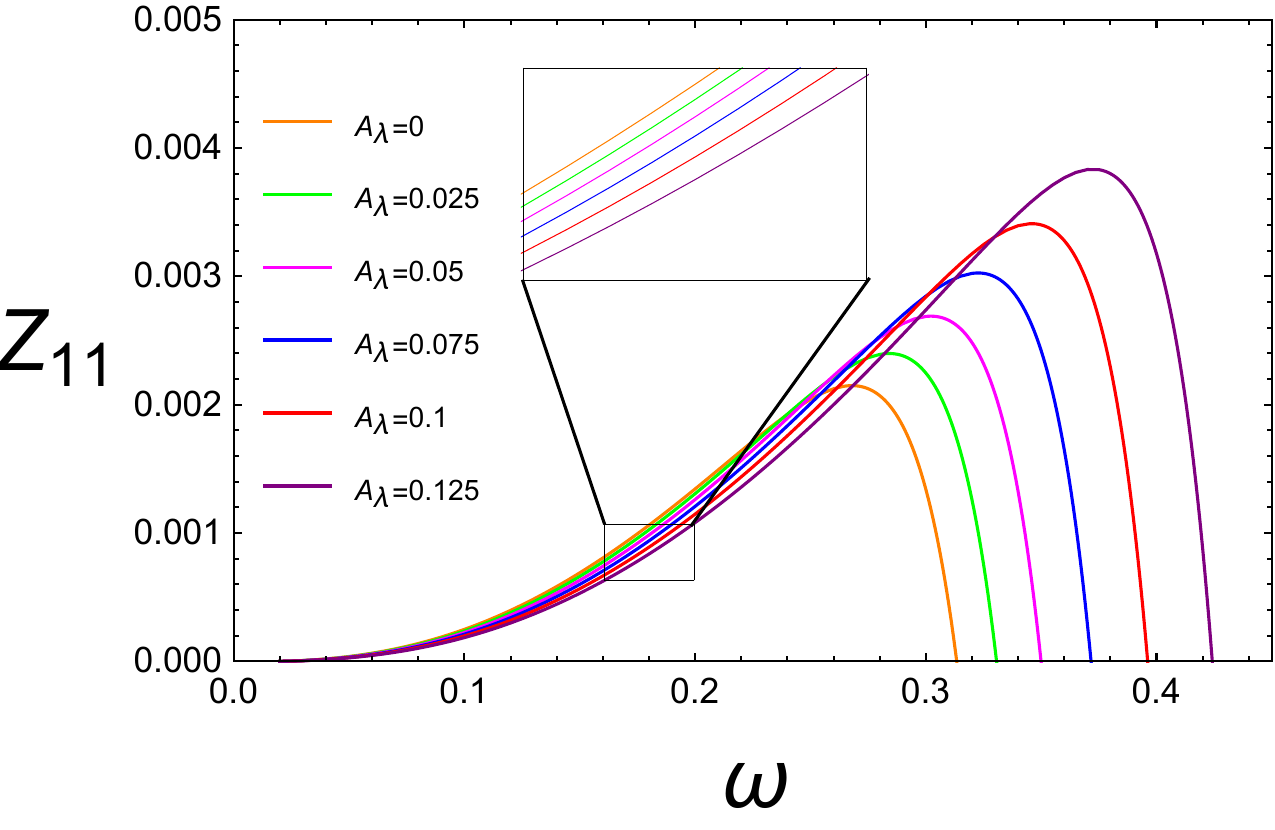}
   		\end{minipage}
   	}
\caption{Amplification factor with respect to the frequency of scalar fields for the mode of $l=1=m$ in rLQGBHs,   where $n=0$, $M_{\rm B}=1$, and $a=0.9$ are set. The left diagram corresponds to the case of massless scalar fields, $\mu=0$, and the right one to the case of massive scalar fields with mass $\mu=0.2$.}
\label{fig:rLQGBH_shooting_superradiance}
\end{figure}

In Fig.~\ref{fig:rLQGBH_shooting_superradiance} we present the relations between the amplification factor and $\omega$ under a varying $A_\lambda$ for a massless incident particle and a massive one with mass $\mu=0.2$, respectively.
In the region where the superradiance effect appears, the energy amplification factor always increases at first and then decreases as the particle frequency $\omega$ grows.
On the boundaries of the region, $\omega=\mu$ and $\omega=m\Omega_\Hor$, see Eq.~(\ref{instability}) with $M_{\rm B}=1$, the amplification factor is always zero.
Moreover, the amplification factor decreases as the particle mass $\mu$ increases, indicating that the superradiance effect is inhibited by particle mass.

The influence of regularization parameter $A_\lambda$ on the amplification factor varies with the particle frequency $\omega$. 
When the particle frequency $\omega$ is low, a small $A_\lambda$ leads to an amplification factor that is larger than that led by a big $A_\lambda$.
The specific situation can be observed in the small picture of  Fig.~\ref{fig:rss-a}, where 
an increase of $A_\lambda$ restrains the superradiance effect in the region of small $\omega$. 
However, as $\omega$ gradually increases, the amplification factor associated with a small $A_\lambda$ is gradually surpassed by the amplification factor associated with a big $A_\lambda$. 
Ultimately, a larger $A_\lambda$ results in a greater peak of amplification factors.
On the whole, an increase of $A_\lambda$ plays the role in promoting the superradiance effect.

%%%%%%%%%%%%%%%%%%%%%%%%%%%%%%%%%%%%%%%%%%%%%%%%%%%%%%%%%%%%%%%%%%
\section{Conclusion}\label{sec:con}

In the present work, we investigate the scalar field perturbations in the background of rLQGBHs. By employing the hyperbolic foliation equation, as expressed in Eq.~\eqref{hyper}, we successfully navigate numerical challenges at the boundaries, enabling us to accurately characterize the time-domain evolution profiles of scalar fields around rLQGBHs. Interestingly, we observe that while $A_\lambda$ does not significantly impact the outburst and late-time tail stages of scalar field evolution, it markedly influences the damping oscillation stage. To further elucidate the effect of $A_\lambda$ on the scalar field evolution, we employ the Prony method, the fourth-order WKB method, and the shooting method to extract QNMs from the damping oscillations. Among these methods, the Prony method emerges as the most precise, while the shooting method is highlighted for its numerical efficiency.

Our exploration reveals that the influence of the scalar field mass $\mu$ on the QNMs is alleviated with an increase of $A_\lambda$. Moreover, we investigate how $A_\lambda$ influences the QNMs by taking different values of angular momentum $a$, uncovering that the effect of $A_\lambda$ on the QNMs varies with the angular momentum. This underscores the significant role played by the regularization parameter in affecting the QNMs of rLQGBHs.

Furthermore, we examine the impact of $A_\lambda$ on the superradiance effect in the background of rLQGBHs, finding that $A_\lambda$ not only broadens the frequency range susceptible to superradiance but also significantly enhances the efficiency of superradiant amplification, particularly when $\omega \sim m\Omega_{\rm H}$.

Based on our comprehensive investigation of QNMs and superradiance, we conclude that the regularization parameter $A_\lambda$ is pivotal in shaping the dynamics of ultralight massive scalar fields in the background of rLQGBHs. Our findings provide novel insights into the behavior of rLQGBHs and set the stage for future inquiries into the dynamical behaviors of rotating LQG wormholes, addressing the challenges posed by their unique boundary conditions. As we chart the course for future research, our work not only enriches the theoretical landscape of black hole physics but also opens new avenues for exploration within the quantum gravity domain.

\section*{Acknowledgments}
Y-GM would like to thank Emmanuele Battista, Stenfan Fredenhagen, and Harold Steinacker for the warm hospitality during his stay at University of Vienna. We would also like to thank Shao-Jun Zhang for his suggestions in the calculation of waveforms. This work was supported in part by the National Natural Science Foundation of China under Grant No.\ 12175108.
%The authors would like to thank  C. Lan for useful discussions.
%This work was supported in part by the National Natural Science Foundation of China under grant Nos. 11675081 and 12175108.
%{\color{red}The authors would like to thank the anonymous referees for the helpful comments that improve this work greatly.}

%\newpage
%\section*{Appendix}
%\appendix

%%%%%%%%%%%%%%%%%%%%%%%%%%%%

%\bibliographystyle{utphys}
\bibliographystyle{unsrturl}
\bibliography{references}
%%%%%%%%%%%%%%%%%%%%%%%%%%%%

\end{document}